\documentclass[preprint,12pt]{elsarticle}
\usepackage[T1]{fontenc}
\usepackage[utf8]{inputenc}
\usepackage{lmodern}
\usepackage{textcomp}
\usepackage{amsmath,amssymb}
\usepackage{graphicx}
\usepackage{booktabs}
\usepackage{multirow}
\usepackage{makecell}
\usepackage{hyperref}

\journal{Urban Climate}
\begin{document}
\begin{frontmatter}
\title{BRUCE (Bayesian Rio Urban Climate Estimator): spatiotemporal interpolation of air temperature from heterogeneous sensor networks for operational heat monitoring}

\author[1,2]{Ayrton Sena Gouveia\corref{cor1}}
\author[1,2]{Rafael Santos de Oliveira}
\author[1,2]{Lucas Menezes-Silva}
\author[2]{Maira Alejandra Moreno Castillo}
\author[1,2]{Felipe de Carvalho Vommaro Marincola}
\author[1,2]{João Henrique de Araújo Morais}
\author[2]{Valéria Saraceni}
\author[2]{Gislani Mateus Oliveira Aguilar}
\author[2]{Caroline Dias Ferreira}
\author[2,3]{Oswaldo Gonçalves Cruz}
\address[1]{Sergio Arouca National School of Public Health, Oswaldo Cruz Foundation (Fiocruz), Rio de Janeiro - RJ, Brazil.}
\address[2]{Rio de Janeiro Municipal Health Secretariat (SMS), Rio de Janeiro - RJ, Brazil}
\address[3]{Program of Scientific Computing, Oswaldo Cruz Foundation (Fiocruz), Rio de Janeiro - RJ, Brazil}
\cortext[cor1]{Corresponding author. E-mail: ayrtonsgouveia19@gmail.com}

\begin{abstract}
Accurate characterization of intra-urban temperature variability is essential for heat-health surveillance, yet official monitoring networks are too sparse to resolve fine-scale gradients, particularly in Global South cities. We present BRUCE, a Bayesian hierarchical spatiotemporal model, fitted with the Integrated Nested Laplace Approximation (INLA) and a Stochastic Partial Differential Equation (SPDE) spatial component, that fuses four heterogeneous temperature sources for Rio de Janeiro, Brazil: official weather stations, low-cost IoT sensors, citizen-science personal weather stations and ERA5-Land reanalysis. The model accounts for class- and station-specific measurement bias and estimates a continuous spatial field that evolves day to day through an autoregressive (AR1) structure. Fitted independently to nine monthly windows (January, June, November; 2023--2025), the elevation coefficient was significant in all windows and of consistent magnitude in eight ($-$1.08 to $-$1.38 \textdegree{}C per SD). In leave-region-out cross-validation against LCZ4r, accuracy was comparable (RMSE differences within station-level bootstrap uncertainty in all three windows), while BRUCE produced predictions at every held-out station, including those where LCZ4r is masked, and preserved finer-scale spatial structure than LCZ4r's discrete zone classification. Applied to 241 primary care territories, BRUCE's posterior exceedance probabilities allowed 96 to be classified with high confidence (model-based classifications), whereas a constant-width interval derived from LCZ4r cross-validation errors supported confident classification of none. The framework offers a statistically principled, operationally viable basis for intra-urban heat monitoring in data-heterogeneous settings.
\end{abstract}
\end{frontmatter}

\section{Introduction}

Urban populations face increasing heat exposure risk as anthropogenic climate change intensifies the frequency, duration, and severity of extreme heat events. The 2025 Lancet Countdown on Health and Climate Change reported that annual heat-related mortality has reached approximately 546,000 deaths globally over the past decade, a 23\% increase since the 1990s, with heat exposure disproportionately affecting urban populations where the built environment amplifies ambient temperatures through the urban heat island (UHI) effect~\cite{romanello_2025_2025}. Such risk is not spatially uniform along a single city, temperature can vary by several degrees over distances of a few hundred meters, driven by heterogeneity in impervious surface cover, vegetation, building density, and topography~\cite{stewart_local_2012}. Accurately characterizing this intra-urban variability is a prerequisite for targeted heat-health surveillance, early-warning systems, and equitable adaptation planning, yet it remains a persistent methodological challenge.

Official meteorological networks, while providing high-quality, long-term observations, are typically too sparse to resolve intra-urban temperature gradients, particularly in cities of the Global South where station density is lower than in well-monitored European or North American cities. The Local Climate Zone (LCZ) classification scheme~\cite{stewart_local_2012} has become a widely adopted framework for structuring urban temperature analysis by land-use archetype. However, LCZ-based approaches alone do not resolve the underlying data-sparsity problem when the observation network itself remains coarse.

The proliferation of low-cost Internet-of-Things (IoT) sensors and citizen-operated personal weather stations (PWS) has been proposed as a means to increase urban observation networks at comparatively low cost~\cite{muller_crowdsourcing_2015,chapman_can_2017,meier_crowdsourcing_2017}. These sources, however, introduce substantial heterogeneity in measurement quality, as such air temperature observations have been repeatedly shown to exhibit systematic warm biases~\cite{bell_how_2015,coney_how_2022}. Existing approaches to this problem have largely relied on deterministic quality-control filtering~\cite{napoly_development_2018,fenner_crowdqcquality-control_2021} followed by treatment of retained observations as equally reliable, an approach that discards potentially informative data and does not formally propagate remaining measurement uncertainty into the resulting spatial product.

Bayesian hierarchical spatiotemporal models offer a principled alternative. In particular, the Stochastic Partial Differential Equation (SPDE) approach to Gaussian random field modeling, implemented via the Integrated Nested Laplace Approximation (INLA), allows continuous spatial (and spatiotemporal) random fields to be estimated at computational cost tractable for large observation networks~\cite{lindgren_explicit_2011}, while natively accommodating multiple observation likelihoods, covariates, and station-level random effects within a single coherent model. This architecture is, in principle, well suited to the sensor-fusion problem described above: rather than filtering heterogeneous sources into a single homogenized dataset, class-specific bias and residual variance can be explicitly modeled and estimated from the data.

Despite this potential, applications that formally integrate official, low-cost IoT, and citizen-science temperature networks within a single Bayesian spatiotemporal framework remain scarce~\cite{fioravanti_interpolating_2023} , particularly for cities with complex coastal-mountainous topography and rapidly evolving observation infrastructure. Rio de Janeiro presents both the need and the opportunity for such an approach as the municipality combines pronounced topographic heterogeneity, an official monitoring network augmented in recent years by a growing low-cost IoT sensor deployment and an established citizen-science PWS community. This study aims to investigate how temperature observations from heterogeneous networks, official weather stations, low-cost IoT sensors, and personal weather stations, can be integrated into a single, statistically coherent spatiotemporal model, and whether this model outperforms simpler interpolation methods currently in operational use.

\section{Methods}

\subsection{Study area}

This study focuses on the municipality of Rio de Janeiro, Brazil, whose official boundary was obtained from the Brazilian Institute of Geography and Statistics (IBGE) via the geobr package~\cite{pereira_geobr_2019}. The municipality presents substantial topographic and land-cover heterogeneity, including coastal areas, dense urban cores, and forested urban massifs (Tijuca, Pedra Branca, and Gericinó), as well as proximity to Guanabara Bay~\cite{solorzano_land_2021}. The city's climate exhibits low seasonal thermal amplitude, with mean maximum temperatures exceeding 27\textdegree{}C and mean minimum temperatures around 23\textdegree{}C between November and April, decreasing to approximately 25\textdegree{}C and 18\textdegree{}C, respectively, from May to September, with the largest recorded amplitude reaching 11.8\textdegree{}C between the warmest and coldest months. This general regime is strongly modulated by local topography and coastal proximity, as the mountainous massifs (Tijuca, Pedra Branca, and Gericinó) generate distinct microclimates, with cooler temperatures and higher precipitation at elevation, while sea breeze circulation moderates temperature and humidity along the coastal lowlands~\cite{neiva_analysis_2017,silva_caracterizacao_2014}.

This topographic complexity has direct methodological implications, because the standard Matérn covariance structure underlying the SPDE approach assumes spatial correlation depends solely on Euclidean distance, implicitly treating the domain as homogeneous and isotropic. This assumption becomes questionable in the presence of physical barriers that restrict atmospheric mixing between nearby locations. This could lead the interpolated field to inappropriately smooth across such barriers, blending microclimates that are, in practice, poorly connected. This motivated the assessment of a physical-barrier extension to the SPDE model, which restricts spatial diffusion across user-specified barrier polygons rather than relying on unmodified Euclidean proximity~\cite{bakka_non-stationary_2019}.

\subsection{Data sources}

Temperature observations were compiled from five station networks spanning three broad classes of data provider: (i) official meteorological networks, comprising automatic stations from the Brazilian National Institute of Meteorology (INMET), the municipal monitoring network (AlertaRio), and airport weather stations (REDEMET); (ii) a low-cost Internet-of-Things (IoT) sensor network (Aurassure), with data available from June 2025; and (iii) citizen-operated personal weather stations reporting through the Weather Underground platform, accessed via its public API (Table 1). The study period spans January 2023 to November 2025.

In addition to the five station networks described above, gridded 2-metre air temperature reanalysis from the ERA5-Land dataset (ECMWF Copernicus Climate Data Store, $\sim$9km horizontal resolution)~\cite{munoz-sabater_era5-land_2021} was incorporated as a fourth observation source. Unlike the station-based sources, ERA5 provides spatially continuous coverage across the state of Rio de Janeiro, extending well beyond the municipal boundary and the spatial extent of the ground-based network. This broader spatial coverage was hypothesized to improve identifiability of the elevation coefficient by providing a wider range of elevation contrast than available within the municipality alone.

\begin{table}[htbp]
\centering
\small
\setlength{\tabcolsep}{4pt}
\caption{Number of distinct stations per network in each of the nine independently-fitted monthly windows. The IoT (Aurassure) network first appears in June 2025, the first window with data from this network; the citizen-science (Weather Underground) and official networks grew steadily over the study period.}
\resizebox{\ifdim\width>\linewidth\linewidth\else\width\fi}{!}{%
\begin{tabular}{lccccc}
\toprule
Window & \makecell{Official\\(INMET / AlertaRio)} & \makecell{Official\\ Airport (REDEMET)} & \makecell{IoT\\(Aurassure)} & \makecell{Citizen-science\\(Weather Underground)} & Total \\
\midrule
\midrule
Jan 2023 & 8 & 5 & 0 & 5 & 18 \\
Jun 2023 & 10 & 5 & 0 & 7 & 22 \\
Nov 2023 & 11 & 5 & 0 & 8 & 24 \\
Jan 2024 & 11 & 5 & 0 & 12 & 28 \\
Jun 2024 & 11 & 5 & 0 & 12 & 28 \\
Nov 2024 & 11 & 5 & 0 & 13 & 29 \\
Jan 2025 & 11 & 5 & 0 & 14 & 30 \\
Jun 2025 & 11 & 5 & 21 & 15 & 52 \\
Nov 2025 & 11 & 5 & 20 & 17 & 53 \\
\bottomrule
\end{tabular}%
}
\end{table}

\subsection{Data quality and control}

Prior to model fitting, observations were subjected to a physical plausibility filter, retaining only daily median and maximum temperature values within the range of 7--45\textdegree{}C. Stations were classified into three origin classes: official, IoT, and citizen-operated. A pair of co-located stations (approximately 248m apart, an official municipal station and a citizen-science Weather Underground station) was found to exhibit a persistent discrepancy in median daily temperature of approximately 1.2\textdegree{}C (SD 1.0\textdegree{}C, n = 83 matched days across three fitting windows), while showing no comparable discrepancy in daily maximum temperature (mean difference $-$0.05\textdegree{}C). This asymmetry motivated the inclusion of the station-level IID effect described below.

During preliminary model development, a class-level fixed effect distinguishing airport weather stations (REDEMET) from other official stations was evaluated, motivated by a suspected warm bias from paved runway surfaces and reduced vegetative cover. Airport stations were found to exhibit heterogeneous bias in opposite directions across stations; however, this heterogeneity is absorbed by the station-level IID effect ($u_{s}$), introduced later in model development, so the class-level term was retained rather than discarded. $\beta_{\mathrm{aeroporto}\  }$ captures the shared class-level offset, while $u_{s}$ absorbs station-specific residual deviations (see S1).

\subsection{Environmental covariates}

The model includes three covariates chosen on physical grounds. Elevation, derived from a digital elevation model (elevatr R package~\cite{hollister_elevatr_2025}) covering the state of Rio de Janeiro, captures the adiabatic temperature gradient and the cooling of the hillslopes and massifs of the city. The built-up (impervious/urban) fraction (0--1), available at municipal extent, represents the thermal inertia, reduced evaporative cooling and anthropogenic heat of impervious surfaces. The Sky View Factor (SVF), the fraction of the sky hemisphere visible from a point (0--1), was obtained at $\sim$137 m resolution from the LCZ4r package's Local Climate Zone classification for Rio de Janeiro. It describes the geometry of the urban canyon: low values, typical of dense and tall building arrangements, reduce the net longwave radiative loss to the sky after sunset and so retain heat, and also shade surfaces during the day. Because the nocturnal effect is expected to be stronger than the daytime one~\cite{stewart_local_2012}, the SVF was included to represent processes relevant to minimum temperature as well as to the daily median. The correlation between SVF and built-up fraction at station locations was moderate (r = $-$0.37), so including both did not raise material multicollinearity concerns.

Elevation and built-up fraction were clipped and masked to the official municipal boundary, ensuring an identical spatial domain for the two rasters. Standardization by z-score was computed from the full distribution of raster cell values within this domain and applied consistently to both station-level extraction and the prediction grid. We also tested whether the built-up fraction would benefit from expansion to state-wide extent, as observed for elevation, using a binary state-wide urban land-cover raster (200 m resolution). Unlike elevation, this did not narrow the coefficient's credible interval (municipal extent: $\beta$ = 0.076, 95\% CrI [$-$0.004, 0.155]; state-wide extent: $\beta$ = 0.075, 95\% CrI [$-$0.012, 0.163]), suggesting that urbanization, unlike elevation, has no spatially continuous regional gradient informative beyond the municipal boundary. An analogous test for the SVF gave the same result, with the credible interval still including zero.

To document this choice, we compared the specification with nested and alternative covariate sets (no covariates, subsets of the three, and the addition of distance to the coast, distance to inland water and building height) in three median windows, using marginal likelihood, WAIC and the station-level log predictive score (Supplementary S1). Elevation was indispensable, and distance to water and building height did not improve the fit. The SVF showed no detectable empirical gain in these windows and was retained on theoretical grounds. Distance to the coast was the only candidate with a consistent coefficient, but its gain in fit was inconsistent across windows. Because the SVF is expected to matter most at night, the same comparison was repeated for the daily minimum temperature in the same three windows (S1): the SVF coefficient was negative in all three (mean $-$0.26 \textdegree{}C per SD) but, as for the median, did not improve model fit.

\subsection{Model structure}

Temperature interpolation was performed using a Bayesian hierarchical spatiotemporal model fitted via the Integrated Nested Laplace Approximation (INLA), with the spatial component represented by a Matérn Gaussian random field approximated through the Stochastic Partial Differential Equation (SPDE) approach~\cite{lindgren_spde_2022}. The linear predictor for observation $i$, at station $s_{i}$ on day $t_{i}$, is given by

\begin{equation}
\begin{aligned}
\eta (s_{i} , t_{i} ) & = \beta_{0} + \beta_{\mathrm{elev}} x_{s_{i} }^{\mathrm{elev}} + \beta_{\mathrm{urb}} x_{s_{i} }^{\mathrm{urb}} + \beta_{\mathrm{aeroporto}} 1 (s_{i} \in \mathrm{aeroporto}\  ) \\ & + \beta_{\mathrm{SVF}} x_{s_{i} }^{\mathrm{SVF}} + \beta_{\mathrm{aura}} 1 (c_{i} = \mathrm{aura}\  ) + \beta_{\mathrm{wunder}} 1 (c_{i} = \mathrm{wunder}\  ) \\ & + u_{s_{i} } + w_{t_{i} } + \xi (s_{i} , t_{i} )
\end{aligned}
\end{equation}

\noindent where $x_{s}^{\mathrm{elev}}$, $x_{s}^{\mathrm{urb}}$ and $x_{s}^{\mathrm{SVF}}$ are standardized elevation, built-up fraction and sky-view factor at station $s$, $c_{i}$ denotes the sensor class of observation $i$, and $u_{s}$, $w_{t}$, $\xi (s , t )$ are random effects described below.

Persistent, sensor-specific bias beyond the class-level average was modeled via an independent and identically distributed (IID) random effect indexed by station,

\begin{equation}
u_{s} \sim \mathcal{N } (0 , \sigma_{u}^{2} ) , \quad \mathrm{i.i.d.\ across\ stations}
\label{eq:iid}
\end{equation}

Four observation likelihoods were specified, reflecting the four data sources: official stations, contributing the reference class with a fixed, tightly specified residual standard deviation ($\sigma_{\mathrm{ofic}} = 0.3$\,\textdegree{}C, reflecting instrument-level measurement precision; S1); IoT (aura) and citizen-science (wunder) stations, each assigned its own independently estimated residual variance informed by a weakly informative PC prior ($P(\sigma > 3) = 0.01$); and ERA5-Land grid cells, assigned a fixed residual standard deviation ($\sigma_{\mathrm{ERA5}} = 0.7$\,\textdegree{}C). This value was set a priori within the range of reported ERA5-Land 2\,m air temperature errors against Brazilian weather stations (RMSE of roughly 0.5--2\,\textdegree{}C for monthly mean temperature, depending on region; e.g., Pernambuco~\cite{araujo_total_2025}) and is intended to represent representativeness error at the $\sim$9\,km grid scale rather than instrument noise. As with the station-class fixed effects, the model formula adapts dynamically to the sources present within a given fitting window.

\begin{equation}
y_{i} \sim \begin{cases} \mathcal{N } (\eta_{i} , \sigma_{\mathrm{ofic}}^{2} ) \quad & \mathrm{if}\  c_{i} = \mathrm{ofic}\   \\ \mathcal{N } (\eta_{i} , \sigma_{\mathrm{aura}}^{2} ) \quad & \mathrm{if}\  c_{i} = \mathrm{aura}\   \\ \mathcal{N } (\eta_{i} , \sigma_{\mathrm{wunder}}^{2} ) \quad & \mathrm{if}\  c_{i} = \mathrm{wunder}\   \\ \mathcal{N } (\eta_{i} , \sigma_{\mathrm{ERA5}}^{2} ) \quad & \mathrm{if}\  c_{i} = \mathrm{ERA5}  \end{cases}
\label{eq:likelihood}
\end{equation}

The spatial field's temporal evolution was modeled via a first-order autoregressive (AR1) structure linking consecutive daily realizations of the Matérn field. During model development, we identified that this spatial AR1 component, when left as the sole temporal element, was forced to jointly represent both genuine local spatial structure and city-wide daily climatic variation shared across all locations (e.g., synoptic-scale weather common to the entire study area on a given day), producing an implausibly large estimated spatial range (on the order of tens of kilometers, exceeding the municipality's physical extent) in a substantial fraction of fitting windows. To resolve this confounding, a separate scalar AR1 term, indexed by day and independent of spatial location,

\begin{equation}
w_{t} = \rho_{w} w_{t - 1 } + \epsilon_{t} , \quad \epsilon_{t} \sim \mathcal{N } (0 , \sigma_{w}^{2} )
\label{eq:ar1}
\end{equation}

\noindent was introduced to capture city-wide daily-level variation, allowing the spatial field $\xi (s , t )$ to represent local structure exclusively. Both AR1 hyperparameters ($\rho_{w}$ and $\sigma_{w}$) were assigned penalized-complexity priors~\cite{simpson_penalising_2017} rather than fixed values: we used PC($u = 0.5$, $\alpha = 0.75$) for $\rho_{w}$ and PC($u = 3$\textdegree{}C, $\alpha = 0.1$) for $\sigma_{w}$, weakly discouraging implausible values while leaving both free to be estimated from the data. Across the nine monthly fitting windows and both temperature statistics, $\sigma_{w}$ was estimated stably throughout (range approximately 1.2--4.3\textdegree{}C).

The choice of a first-order autoregressive structure for the daily-level term was formally assessed via autocorrelation (ACF) and partial autocorrelation (PACF) analysis of city-mean daily temperature, computed independently across five seasonal windows (spanning both summer and winter) for both median and maximum daily temperature (ten combinations total). In all ten combinations, only the first-lag partial autocorrelation was consistently significant, with no repeated higher-order or cyclical pattern across windows, supporting AR1 as the appropriate structure at the daily scale, following standard Box-Jenkins order-identification practice.

\subsection{Prior specification}

Hyperparameters throughout the model were assigned through penalized-complexity (PC) priors. The Matérn field's range and standard deviation were assigned PC($u = 15{,}000$m, $\alpha = 0.5$) and PC($u = 8$\textdegree{}C, $\alpha = 0.01$) priors, respectively. An empirical semivariogram of covariate-adjusted station residuals showed no clearly resolvable sill within the distances sampled by the network, consistent with the identifiability limitation discussed in the Limitations rather than providing an independent point estimate for this parameter.

The spatial field's own AR1 group correlation (GroupRho) was assigned the same PC($u = 0.5$, $\alpha = 0.75$) prior as $\rho_{w}$, for consistency. This choice was not miscalibrated relative to the data as among freely-converged windows, empirical GroupRho estimates averaged 0.70 (median temperature) and 0.40 (maximum temperature), both well within the prior's plausible mass. In windows where this parameter instead collapsed to the boundary of its parameter space (GroupRho $\rightarrow$ 1), it was fixed at 0.55 --- a value between the two statistics' own empirically-derived references --- and validated via a sensitivity analysis confirming this fixed value did not materially alter the elevation coefficient ($\beta_{\mathrm{elev}}$) relative to refitting with each statistic's own reference value instead (S1). The station-level effect ($u_{s}$) was assigned a weakly-informative PC($u = 1$\textdegree{}C, $\alpha = 0.01$) prior, penalizing implausibly large per-station bias while allowing it to be estimated freely.

Observation-level noise was fixed a priori for official stations ($\sigma_{\mathrm{ofic}} = 0.3$\textdegree{}C) and ERA5-Land pseudo-observations ($\sigma_{\mathrm{ERA5}} = 0.7$\textdegree{}C), and assigned PC($u = 3$\textdegree{}C, $\alpha = 0.01$) priors for the low-cost IoT (Aurassure) and citizen-science (Weather Underground) networks. The official-station value was independently validated against detrended overnight station residuals (0.356\textdegree{}C; S1). In a full-model sensitivity test where $\sigma_{\mathrm{ofic}}$ was instead estimated freely, the freely-estimated Aurassure and Weather Underground (0.686\textdegree{}C and 0.576\textdegree{}C, respectively) remained comfortably below the PC prior's implausibility threshold.

\subsection{Temporal windowing strategy}

Independent monthly fitting was adopted for three reasons. First, jointly fitting a single spatiotemporal AR1 model across the full 39-month study period is computationally infeasible given the joint precision structure linking the spatial field across all days. Second, the composition of the observation network itself varied substantially over the study period --- most notably, the IoT (Aurassure) network was entirely absent prior to June 2025 --- such that a single pooled model would conflate distinct data-generating regimes under a common set of hyperparameters. Lastly, independent refitting across nine windows doubles as an internal robustness check: rather than assuming the model's central findings (e.g., the elevation effect) generalize from a single fit, this design allows their stability to be directly assessed across independently-estimated windows spanning different seasons and network configurations.

Spatial range, spatial-temporal correlation (GroupRho), and daily-level AR1 correlation ($\rho_{w}$) were compared across the nine windows (January, June, November; 2023--2025), separately for median and maximum temperature. For maximum temperature, only two of the nine windows (June and November 2025) converged to a stable GroupRho estimate (0.397 and 0.407, respectively); the remaining seven (January, June, and November 2023; January, June, and November 2024; January 2025) collapsed to the boundary of the parameter space (GroupRho $\rightarrow$ 1) and were fixed at 0.55, as described in Prior specification. For median temperature, six windows converged to GroupRho estimates between 0.64 and 0.80, and the remaining three (January, June, and November 2024) exhibited the same boundary collapse documented above for maximum temperature and were stabilized using the same fixing approach (fixed at 0.55; see Prior specification). Among these six converged windows, range and $\rho_{w}$ were most extreme in November 2025 (37.0 km; $\rho_{w}$ = 0.49) --- markedly more so than in June 2025 (15.0 km; $\rho_{w}$ = 0.74), a similarly-sized deployment of the low-cost IoT network (Aurassure: 20 vs. 21 stations), arguing against network composition as the primary driver. This window also contains 8 November 2025, a day with unusually large validation errors (S1), which may contribute to its reduced day-to-day persistence; excluding that day, however, changed the November validation RMSE only marginally (S1).

One window (June 2023) exhibited a notably smaller spatial range (2.3 km) combined with larger field magnitude (Stdev $\approx$ 1.3\textdegree{}C) than the remaining eight windows, alongside markedly elevated relative parameter uncertainty (25\% for range, 20\% for Stdev, versus $\sim$10--15\% typical of the other windows) --- a pattern more consistent with weak parameter identifiability in this window than with a genuine seasonal effect, reinforcing the range-identifiability limitation discussed above.

\subsection{Model validation and comparison}

Model performance was assessed via a spatial block cross-validation protocol (leave-region-out) to specifically evaluate the model's capacity for spatial extrapolation into areas without nearby ground-truth observations, addressing a known limitation of randomly partitioned cross-validation in the presence of spatial autocorrelation~\cite{roberts_cross-validation_2017}. Stations were partitioned into five blocks corresponding to the municipality's official Planning Areas (Áreas de Planejamento), a real administrative geography rather than an arbitrary geometric partition, avoiding the risk of a partition scheme that inadvertently isolates covariate outliers (e.g., a single high built-up-fraction airport station) into a training-poor block. Minimum nearest-neighbor distance between held-out and retained stations across partitions had a median of 8,432m (range: 3.2--23.5km, across all region $\times$ station $\times$ month combinations in the three fitting windows used for this comparison). This distance exceeded the window's own estimated spatial range only in January 2025 (range: 7.7km); in June and November 2025, the estimated range (15.0km and 37.0km, respectively) exceeded the inter-station spacing achieved by the administrative partition. This asymmetry does not indicate a partition-specific leakage artifact --- the same administrative blocks were used in all three months --- but rather reflects the broader range-identifiability limitation discussed in Limitations, which is itself most pronounced in these same large-range windows.

The same partition was applied to the BRUCE model described above, and interpolation via the LCZ4r package. For the BRUCE model, held-out regions were excluded from all station-based likelihoods (official, IoT, citizen-science) while ERA5 background coverage was retained, mirroring the availability of background covariate information (elevation, built-up fraction) to the comparison methods.

For test stations of the airport class, predictions included the class-level airport effect ($\beta_{\mathrm{aeroporto}}$) estimated in the corresponding fold from the airport stations retained in training; the station-specific effect $u_{s}$ of a held-out station is unavailable by construction and was not added. In one fold (region 5, January 2025), GroupRho collapsed to unity under free estimation, and the fold was refitted with GroupRho fixed at 0.77, the value estimated by the full-data fit of that window; in the other 14 folds GroupRho was estimated freely.

Methods were compared on the common set of held-out station-days for which both produced a valid estimate. Accuracy was summarized by RMSE, and uncertainty in the RMSE difference (BRUCE minus LCZ4r) was quantified by a nonparametric bootstrap that resamples held-out stations (2,000 replicates), which respects the clustering of daily observations within stations. The distribution of absolute errors was further described by its median and upper quantiles, with paired Wilcoxon signed-rank tests on absolute error as a secondary analysis; because these tests treat observations as independent, their p-values are likely optimistic.

Comparisons against LCZ4r should be interpreted with the following caveat: BRUCE's temporal AR1 structure allows each day's prediction to borrow statistical strength from the full fitting window, whereas LCZ4r, used here in its standard day-by-day mode (no native temporal pooling), predicts each day independently from same-day observations alone. This asymmetry reflects each method's typical operational use rather than an artificially constructed advantage, but should be considered when interpreting relative performance.

Predictive uncertainty was propagated via posterior sampling using INLA (500 samples), drawn jointly from the fitted model's fixed effects, spatial field, and daily-level term. For each sample, the full spatial prediction was reconstructed and pixel-level summary statistics (mean, standard deviation, 95\% credible interval) were computed empirically across samples, rather than relying on posterior means alone. Predictive calibration was further assessed via the continuous ranked probability score (CRPS) at leave-region-out test locations, computed from 200 joint posterior draws of the latent temperature field (fixed effects, daily term and spatial field) at each held-out station. This predictive distribution describes the noise-free temperature at the location and therefore excludes the station-specific effect and the measurement error of a new sensor (and, for airport stations, the class-level airport effect); CRPS is reported for BRUCE only, since LCZ4r does not provide a predictive distribution.

\subsection{Population heat exposure across primary care territories}

To illustrate the implications of interpolation choice for heat-health surveillance, we assessed population exposure on the hottest day of the study period (18 November 2023). Residential population was taken from a 1 km gridded population product for 2024~\cite{bondarenko_censusprojection-disaggregated_2020}, distributed to 500 m grid cells in proportion to overlapping area, and allocated to the 241 primary care (APS) service territories of the municipality, rescaling each territory's total to its registered population. For each posterior sample, we computed the number of residents in grid cells exceeding daily median temperature thresholds of 30, 32 and 34\textdegree{}C, and the population-weighted mean temperature of each territory, posterior means, and 95\% credible intervals were obtained across samples. Territories were classified as above or below the 32\textdegree{}C threshold when the posterior probability of exceedance was greater than 0.9 or less than 0.1, respectively, and as uncertain otherwise.

For comparison, LCZ4r was applied to the same station data for the same day at 500 m resolution, matching the BRUCE grid, and aggregated to territories in the same way. Because LCZ4r does not provide a spatially varying measure of prediction uncertainty, an empirical interval was derived from same-day leave-one-out cross-validation errors of LCZ4r exceedance. Probabilities for each territory were computed from this error distribution, which implies an interval of constant width across the city.

\subsection{Software and computational implementation}

All analyses were performed on a Linux Mint 22.3 workstation equipped with 20 logical CPUs (10 physical cores with hyperthreading) Intel(R) Core i9-10900X and 64 GB of RAM. Model fitting used R version 4.6.1~\cite{r_core_team_r_2026} and the R-INLA package version 26.08.07~\cite{lindgren_bayesian_2015}, sf and terra for spatial data handling~\cite{pebesma_spatial_2023,hijmans_terra_2020}, gstat~\cite{pebesma_multivariable_2004} for exploratory geostatistical analysis, geobr~\cite{pereira_geobr_2019} for administrative boundaries, and DuckDB~\cite{raasveldt_duckdb_2019} as the underlying data storage backend. We adopted a Gaussian approximation to the latent field with a central composite design (CCD) integration strategy for the hyperparameters (strategy = ``gaussian'', int.strategy = ``ccd'') throughout. Monthly windows were fitted with limited parallelism (2 concurrent processes) to remain within memory constraints identified during preliminary testing, and need on average $\sim$34 minutes to fit.

\section{Results}

\subsection{Relative contribution of covariates and the spatial field}

To characterize the relative contribution of the covariate-driven mean structure versus the residual spatial field to the final predicted temperature surface, we decomposed the spatially-varying part of the linear predictor into two components --- $\beta_{\mathrm{elev}} x_{s}^{\mathrm{elev}} + \beta_{\mathrm{urb}} x_{s}^{\mathrm{urb}}$ and $\xi (s , t )$ --- evaluated on the same 500m prediction grid for a representative day (15 January 2023, median temperature model).

The covariate-driven component exhibited an amplitude of 4.3\textdegree{}C across the grid (range: $-$2.6\textdegree{}C to 1.7\textdegree{}C; SD = 0.71\textdegree{}C), reflecting the smooth, large-scale gradient associated with elevation and built-up fraction. The spatial field component exhibited a smaller amplitude (2.8\textdegree{}C; range: $-$0.8\textdegree{}C to 2.0\textdegree{}C) and standard deviation (0.47\textdegree{}C) than the covariate component, but its magnitude was strongly associated with proximity to the station network: |$\xi (s , t )$| averaged 0.6--0.7\textdegree{}C within 2km of the nearest contributing station, falling to 0.32\textdegree{}C at 5--10km and 0.15\textdegree{}C beyond 10km (Spearman $\rho$ = $-$0.40, p < 0.001), confirming that the field's contribution is concentrated near observed locations and decays with distance rather than being distributed uniformly across the domain. Decomposing total spatial variance across the grid, the covariate component accounted for approximately 70\% of variance, with the residual field accounting for the remaining 30\%.

To assess whether this decomposition generalizes beyond a single day, we repeated it across all nine independently-fitted monthly windows (median temperature). The spatial field accounted for 25.0\% of variance on average (range 6.4--42.9\%), and its magnitude decayed with distance to the nearest station in all nine windows (mean Spearman $\rho$ = $-$0.31, range $-$0.55 to $-$0.13). Within each independently-fitted window, the field's relative contribution to total variance increased on days of greater city-wide thermal amplitude as the Spearman correlation between daily amplitude and the field's share of variance was positive in all 18 window$\times$statistic combinations (mean $\rho$ = 0.63, range 0.34--0.86) and significant at p < 0.05 in 17 of 18; the single exception (median, November 2023) was still positive ($\rho$ = 0.34, p = 0.064).

This pattern indicates that elevation and built-up fraction capture the dominant large-scale spatial structure of urban temperature in the study area, while the spatial random field contributes substantial but spatially-localized corrections, reaching up to approximately 2\textdegree{}C within 500m of individual stations, that decay with distance, reflecting both genuine fine-scale microclimatic variation and the network's limited spatial resolution for resolving structure beyond station-adjacent areas.

Figure 1 illustrates this pattern across four representative days spanning different months and magnitudes of spatial contrast (15 Nov 2023, 17 Jan 2024, 17 Jan 2025, 16 Jun 2025): (a) the total predicted surface, (b) the covariate-driven component --- smooth and constant within each month --- and (c) the residual spatial field, concentrated near stations and varying in magnitude from 20\% to 52\% of local variance across the four days shown.

\begin{figure}[htbp]
\centering
\includegraphics[width=\linewidth]{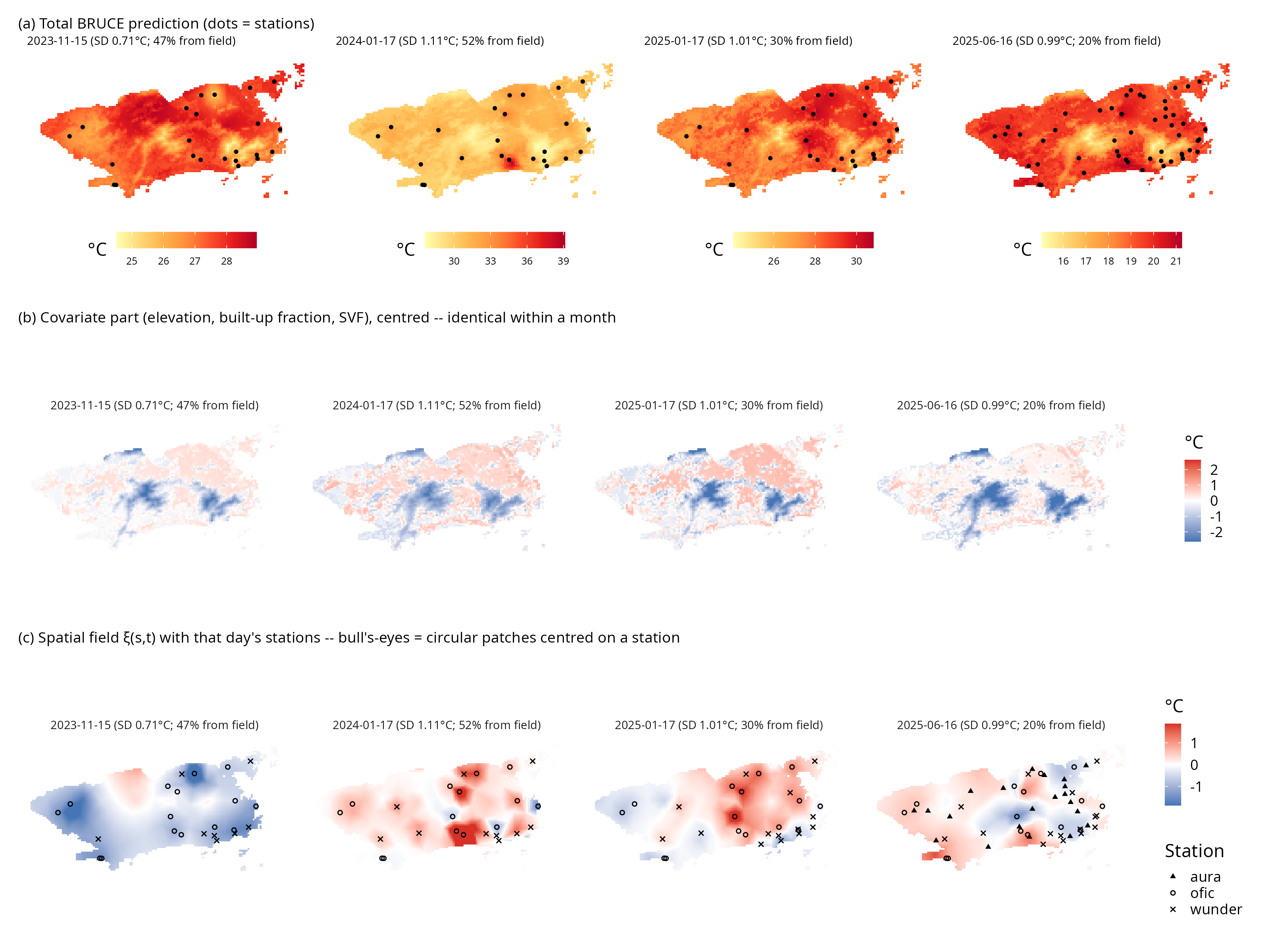}
\caption{Decomposition of the predicted temperature surface (median model) into its covariate-driven and spatial-field components, for four days spanning different months and magnitudes of spatial contrast: 15 November 2023, 17 January 2024, 17 January 2025, and 16 June 2025 (ranked 185th, 6th, 52nd, and 60th, respectively, by total spatial contrast among the 242 days across the nine fitting windows). (a) Total predicted surface, shown with an independent colour scale per day to highlight each day's own spatial pattern. (b) Covariate-driven component (elevation, built-up fraction and SVF). (c) Residual spatial field $\xi$(s,t), shown on a shared colour scale, with that day's contributing stations overlaid. The spatial field accounted for 47\%, 52\%, 30\%, and 20\% of total spatial variance on these four days, respectively --- a range consistent with that observed across all nine independently-fitted monthly windows.}
\end{figure}

\subsection{Stability of the elevation effect across independent fitting windows}

The elevation coefficient remained significant across all nine windows (95\% credible intervals never crossing zero in any window; Table 2 and 3) and of broadly consistent magnitude in eight of nine windows ($\beta_{\mathrm{elev}\  }$ range: $-$1.08 to $-$1.38\textdegree{}C per SD of elevation). One window (November 2025) exhibited a comparatively weaker elevation coefficient ($-$0.79\textdegree{}C). In contrast, the built-up fraction coefficient exhibited significance in every June ($\beta_{\mathrm{urb}\  }$ range 0.14--0.19) and January window ($\beta_{\mathrm{urb}\  }$ range 0.10--0.14), but no November window ($\beta_{\mathrm{urb}\  }$ range 0.04--0.08, all credible intervals crossing zero).  This November-specific attenuation, rather than a clean winter/summer gradient, may instead reflect November's transitional position between the dry and wet seasons in Rio de Janeiro's climate. The airport class effect was positive in all nine windows (posterior means $+$1.37 to $+$2.32\,\textdegree{}C), with credible intervals excluding zero in every one.

\begin{table}[htbp]
\centering
\small
\setlength{\tabcolsep}{4pt}
\caption{Fixed-effect coefficients (standardized)
  across nine independently-fitted monthly windows (median temperature). Elevation remained significant in every window. Built-up fraction reached significance in every June window but no November window. Sky View Factor did not reach significance in any window. The airport class effect was positive, with a credible interval excluding zero, in every window. }
\label{tab:coefs-9-months}
\resizebox{\ifdim\width>\linewidth\linewidth\else\width\fi}{!}{%
\begin{tabular}{lcccccccc}
\toprule
Window & \multicolumn{2}{c}{$\beta_{\mathrm{elev}}$} & \multicolumn{2}{c}{$\beta_{\mathrm{urb}}$} & \multicolumn{2}{c}{$\beta_{\mathrm{svf}}$} & \multicolumn{2}{c}{$\beta_{\mathrm{aeroporto}}$} \\
\cmidrule(lr){2-3}\cmidrule(lr){4-5}\cmidrule(lr){6-7}\cmidrule(lr){8-9}
 & Mean & 95\% CrI & Mean & 95\% CrI & Mean & 95\% CrI & Mean & 95\% CrI \\
\midrule
Jan 2023 & $-$1.28 & ($-$1.37, $-$1.19) & 0.095 & (0.01, 0.18) & $-$0.26 & ($-$0.87, 0.36) & 2.19 & (1.02, 3.35) \\
Jun 2023 & $-$1.30 & ($-$1.40, $-$1.20) & 0.187 & (0.10, 0.28) & 0.34 & ($-$0.37, 1.06) & 1.91 & (0.49, 3.32) \\
Nov 2023 & $-$1.08 & ($-$1.18, $-$0.99) & 0.075 & ($-$0.00, 0.15) & $-$0.11 & ($-$0.57, 0.35) & 1.37 & (0.43, 2.30) \\
Jan 2024 & $-$1.34 & ($-$1.41, $-$1.27) & 0.142 & (0.08, 0.20) & $-$0.05 & ($-$0.43, 0.33) & 1.61 & (0.74, 2.48) \\
Jun 2024 & $-$1.17 & ($-$1.26, $-$1.09) & 0.156 & (0.08, 0.24) & 0.15 & ($-$0.62, 0.92) & 2.32 & (0.55, 4.07) \\
Nov 2024 & $-$1.19 & ($-$1.26, $-$1.12) & 0.035 & ($-$0.03, 0.10) & $-$0.44 & ($-$1.04, 0.16) & 1.99 & (0.53, 3.45) \\
Jan 2025 & $-$1.38 & ($-$1.48, $-$1.29) & 0.131 & (0.05, 0.21) & $-$0.32 & ($-$0.70, 0.07) & 1.95 & (1.01, 2.88) \\
Jun 2025 & $-$1.13 & ($-$1.24, $-$1.02) & 0.143 & (0.06, 0.23) & 0.24 & ($-$0.09, 0.57) & 2.02 & (0.93, 3.10) \\
Nov 2025 & $-$0.79 & ($-$0.90, $-$0.68) & 0.055 & ($-$0.02, 0.13) & $-$0.03 & ($-$0.29, 0.23) & 1.72 & (0.86, 2.59) \\
\bottomrule
\end{tabular}%
}
\end{table}

\subsection{Robustness to daily temperature statistic (maximum vs. median)}

For the maximum-temperature model, GroupRho exhibited the same collapse-toward-unity pattern documented for a subset of median-temperature windows, occurring more frequently given max temperature's stronger diurnal forcing. GroupRho was fixed at 0.55 for the seven affected windows, as described in Prior specification. A sensitivity analysis refitting one median window (January 2024) and one maximum window (November 2023) with GroupRho fixed at the empirical value of each statistic's freely converged windows (0.70 and 0.40, respectively) instead of 0.55 changed the elevation coefficient by 0.01 and 0.03 \textdegree{}C (S1), well within one posterior standard deviation; this pragmatic fix therefore does not materially bias the elevation effect.

\begin{table}[htbp]
\centering
\small
\setlength{\tabcolsep}{4pt}
\caption{Fixed-effect coefficients (standardized) across nine independently-fitted monthly windows (maximum temperature). GroupRho was fixed at 0.55 for seven windows exhibiting the collapse-toward-unity pattern (S1); the remaining two converged without intervention. Elevation remained significant in every window, with magnitudes close to those observed for median temperature (Table 2). The airport class effect was positive in every window, but its credible interval excluded zero in only three (January and June 2023, June 2025).}
\label{tab:coefs-max-9-months}
\resizebox{\ifdim\width>\linewidth\linewidth\else\width\fi}{!}{%
\begin{tabular}{lcccccccc}
\toprule
Window & \multicolumn{2}{c}{$\beta_{\mathrm{elev}}$} & \multicolumn{2}{c}{$\beta_{\mathrm{urb}}$} & \multicolumn{2}{c}{$\beta_{\mathrm{svf}}$} & \multicolumn{2}{c}{$\beta_{\mathrm{aeroporto}}$} \\
\cmidrule(lr){2-3}\cmidrule(lr){4-5}\cmidrule(lr){6-7}\cmidrule(lr){8-9}
 & Mean & 95\% CrI & Mean & 95\% CrI & Mean & 95\% CrI & Mean & 95\% CrI \\
\midrule
Jan 2023 & $-$0.98 & ($-$1.09, $-$0.87) & 0.036 & ($-$0.05, 0.12) & $-$0.43 & ($-$1.55, 0.69) & 2.25 & (0.18, 4.31) \\
Jun 2023 & $-$0.82 & ($-$0.91, $-$0.72) & $-$0.008 & ($-$0.08, 0.06) & 0.14 & ($-$0.85, 1.13) & 2.10 & (0.15, 4.04) \\
Nov 2023 & $-$0.94 & ($-$1.08, $-$0.80) & 0.011 & ($-$0.10, 0.12) & 0.09 & ($-$0.62, 0.80) & 0.73 & ($-$0.70, 2.15) \\
Jan 2024 & $-$0.67 & ($-$0.77, $-$0.56) & 0.012 & ($-$0.06, 0.09) & 0.48 & ($-$0.25, 1.21) & 1.14 & ($-$0.52, 2.79) \\
Jun 2024 & $-$0.79 & ($-$0.89, $-$0.70) & $-$0.003 & ($-$0.07, 0.07) & 0.76 & ($-$0.29, 1.81) & 1.57 & ($-$0.80, 3.92) \\
Nov 2024 & $-$0.70 & ($-$0.83, $-$0.57) & 0.005 & ($-$0.09, 0.10) & 0.26 & ($-$0.53, 1.05) & 1.39 & ($-$0.49, 3.27) \\
Jan 2025 & $-$0.80 & ($-$0.92, $-$0.69) & 0.022 & ($-$0.06, 0.10) & 0.47 & ($-$0.37, 1.31) & 1.40 & ($-$0.62, 3.42) \\
Jun 2025 & $-$1.28 & ($-$1.36, $-$1.20) & 0.041 & ($-$0.04, 0.12) & 0.15 & ($-$0.25, 0.54) & 1.41 & (0.09, 2.72) \\
Nov 2025 & $-$0.71 & ($-$0.82, $-$0.60) & 0.014 & ($-$0.06, 0.09) & 0.06 & ($-$0.39, 0.51) & 0.61 & ($-$0.89, 2.10) \\
\bottomrule
\end{tabular}%
}
\end{table}

The elevation coefficient for maximum temperature ($\beta_{\mathrm{elev}\  }$ range: $-$0.67 to $-$1.28\textdegree{}C per SD) closely matched the magnitude observed for median temperature ($-$0.79 to $-$1.38\textdegree{}C; Table 2), providing independent confirmation of this study's central finding across two distinct temperature statistics. The airport class effect was again positive in all nine windows ($+$0.61 to $+$2.25\,\textdegree{}C), but its credible interval excluded zero in only three of them (January and June 2023, June 2025), indicating weaker evidence for a shared airport offset in the daily maximum than in the daily median.

\subsection{Comparison against reference interpolation methods}

Because LCZ4r cannot produce a valid estimate at some stations (coastal/water Local Climate Zone masking: all of Planning Area 1 and several stations in Planning Areas 3--5), methods were compared on the common set of held-out station-days (n = 302, 285 and 276 in January, June and November 2025). Leave-region-out RMSE was 1.90 vs 2.21 \textdegree{}C in January, 2.13 vs 2.08 \textdegree{}C in June and 1.76 vs 1.83 \textdegree{}C in November for BRUCE and LCZ4r, respectively. Station-level bootstrap intervals for the RMSE difference (BRUCE minus LCZ4r) included zero in all three windows (January: $-$0.31 \textdegree{}C, 95\% CI $-$0.87 to 0.27; June: 0.05 \textdegree{}C, $-$0.77 to 0.94; November: $-$0.07 \textdegree{}C, $-$0.60 to 0.52), indicating comparable accuracy. Excluding airport stations (n = 210, 196 and 186), RMSE was 1.69 vs 1.88, 2.12 vs 2.12 and 1.52 vs 1.71 \textdegree{}C.

Error distributions differed in shape. The median absolute error was lower for LCZ4r than for BRUCE in June and November (BRUCE vs LCZ4r: 1.81 vs 1.18 and 1.28 vs 0.91\,\textdegree{}C; paired Wilcoxon $p = 0.008$ and $0.042$, observation-level and likely optimistic given within-station correlation); in January the difference was not significant (1.36 vs 1.11\,\textdegree{}C, $p = 0.63$). In contrast, BRUCE had a lower 90th-percentile absolute error in all three windows (BRUCE vs LCZ4r: 3.03 vs 3.88, 3.32 vs 3.58 and 2.74 vs 3.37\,\textdegree{}C) and similar maxima (6.1--6.3 vs 5.3--6.7\,\textdegree{}C), consistent with the comparable RMSE.

CRPS for BRUCE (latent-field predictive distribution; see Methods) was 1.16 \textdegree{}C in January 2025 (n = 457), 1.65 \textdegree{}C in June 2025 (n = 431) and 1.16 \textdegree{}C in November 2025 (n = 426; overall mean 1.32 \textdegree{}C), consistent with the RMSE-based comparison in identifying June as the most challenging fitting window. CRPS is reported for BRUCE only, as LCZ4r's deterministic variogram-based interpolation does not produce a comparable predictive distribution.

At the individual region level, restricted to the 12 of 15 region $\times$ month configurations in which both methods produced a valid estimate, BRUCE had the lowest RMSE in 6 and LCZ4r in 6 (region- and month-level detail in S1): BRUCE was clearly better in Planning Area 2 in all three months, whereas LCZ4r was better in Planning Area 3 (where it is evaluated at a single station) and, in June and November, in Planning Area 5. BRUCE was the only method producing a valid estimate in Planning Area 1 in all three months, and over the full network its RMSE was 1.82, 2.20 and 1.66 \textdegree{}C (n = 457, 431, 426).

While aggregate accuracy was comparable between methods, the two produce visually distinct spatial surfaces. Figure 2 illustrates this using each method's full-network prediction for 18 November 2023, the hottest day of the study period.

\begin{figure}[htbp]
\centering
\includegraphics[width=\linewidth]{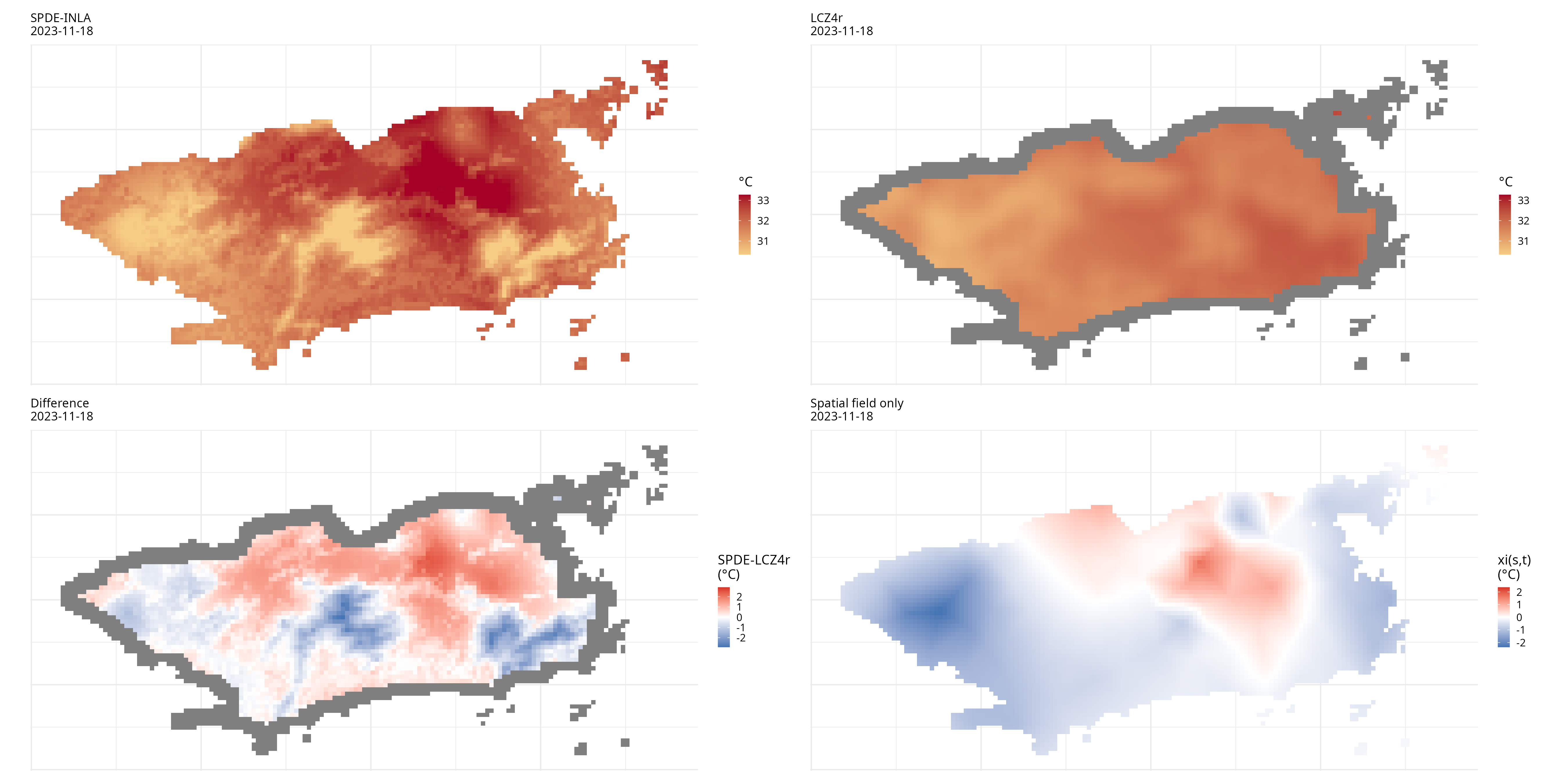}
\caption{Spatial comparison of BRUCE and LCZ4r full-network predictions for 18 November 2023, the hottest day of the study period. (Top-Left) BRUCE predicted surface. (Top-Right) LCZ4r predicted surface. (Bottom-Left) Difference (BRUCE minus LCZ4r). (Bottom-Right) Effect of spatial field.}
\end{figure}

The two methods differ markedly in visual character despite comparable aggregate predictive accuracy: LCZ4r's variogram-based interpolation produces visually coherent, smooth surfaces by design, while BRUCE's shorter estimated spatial range preserves substantial local heterogeneity, including local disagreement between closely-spaced low-cost sensors. Two lines of evidence, detailed in S1, indicate that this heterogeneity reflects genuine local signal rather than noise. First, BRUCE's predictions correlated more strongly with held-out observations than LCZ4r's in all three windows (pooled r = 0.84, 0.79 and 0.69 vs 0.69, 0.72 and 0.60). After removing each day's mean to isolate spatial variation, within-day correlations were 0.57, 0.54 and 0.41 for BRUCE vs $-$0.02, 0.12 and 0.07 for LCZ4r, and remained higher for BRUCE when airport stations were excluded (0.76, 0.69 and 0.38 vs 0.06, 0.23 and 0.12). Second, the station-level effect ($u_{s}$) of four of the five airport stations kept a consistent sign and magnitude across the 12 leave-region-out refits in which each was retained in training (SBAF +0.9, SBJR +1.5, SBGL $-$1.1 and SBSC $-$1.1 \textdegree{}C on average; SBRJ $\approx$ 0).

\subsection{Predictive Uncertainty}

Pixel-level standard deviation for 18 November 2023 (the hottest day of the study period; November 2023 fitting window) ranged from 0.28\textdegree{}C to 0.82\textdegree{}C (mean 0.57\textdegree{}C across n = 5,222 grid pixels), broadly consistent in magnitude with the estimated spatial field standard deviation for this window (0.75\textdegree{}C), confirming that the propagated uncertainty reflects genuine model-based variability rather than an artifact of the sampling procedure.

\subsection{Station-level bias correction}

Beyond class-level bias correction ($\beta_{\mathrm{aura}\  }$, $\beta_{\mathrm{wunder}\  }$), the station-level IID effect ($u_{s}$) identified substantial individual sensor bias not captured by class membership alone. Twenty-nine of 52 stations (55.8\%) exhibited a credible non-zero bias (95\% credible interval excluding zero; Figure 3), ranging from $-$2.85\textdegree{}C to +3.67\textdegree{}C. This station-specific correction capability, native to the hierarchical Bayesian framework, has no counterpart in deterministic interpolation methods such as LCZ4r, which treat all input observations as equally reliable.

The proportion of stations exhibiting credible individual bias was comparable between official and IoT sources (official: 7/15, 46.7\%; IoT: 9/20, 45.0\%) but markedly higher among citizen-science stations (wunder: 13/17, 76.5\%), indicating that within-class site-specific deviation is not confined to low-cost sources, though the citizen-science network carries a higher overall rate. For official stations specifically, this individual effect is more plausibly attributable to genuine unmodeled site-level microclimatic variation (e.g., local shading, sheltering) than to instrument miscalibration: the same seven official stations exhibited a credible bias of consistent sign in an independent refitting under a different network configuration (November 2023), with all remaining eight official stations credible in neither window --- a reproducibility pattern unlikely to arise from measurement noise.

\begin{figure}[htbp]
\centering
\includegraphics[width=\linewidth]{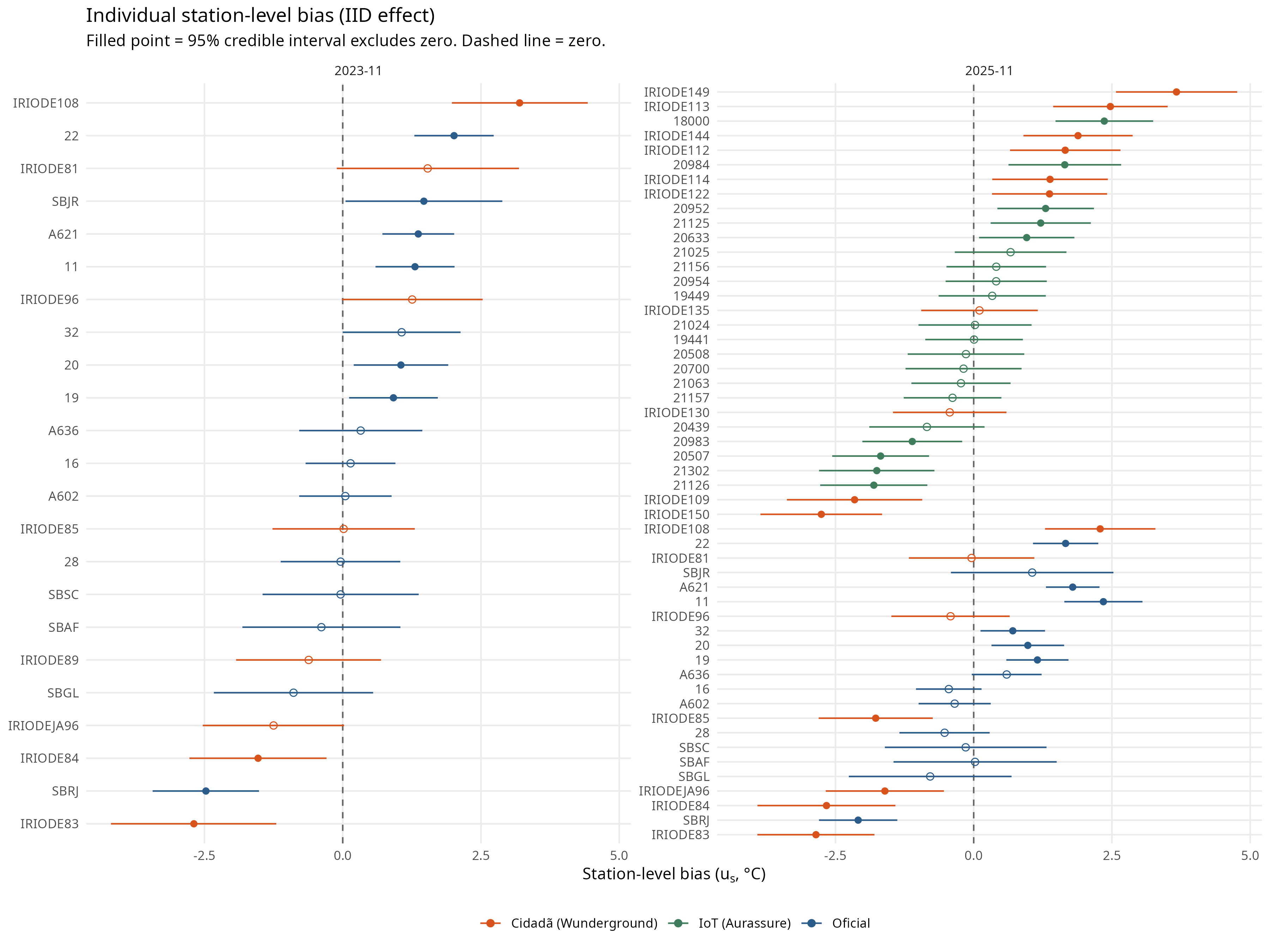}
\caption{Estimated station-level bias (IID effect, $u_{s}$) for all stations, November 2023 (n=23) and November 2025 (n=52). Error bars show 95\% credible intervals; intervals not crossing zero (dashed line) indicate statistically credible individual bias beyond class-level correction. Stations are ordered by estimated bias within each panel.}
\end{figure}

\subsection{Population heat exposure}

\begin{figure}[htbp]
\centering
\includegraphics[width=\linewidth]{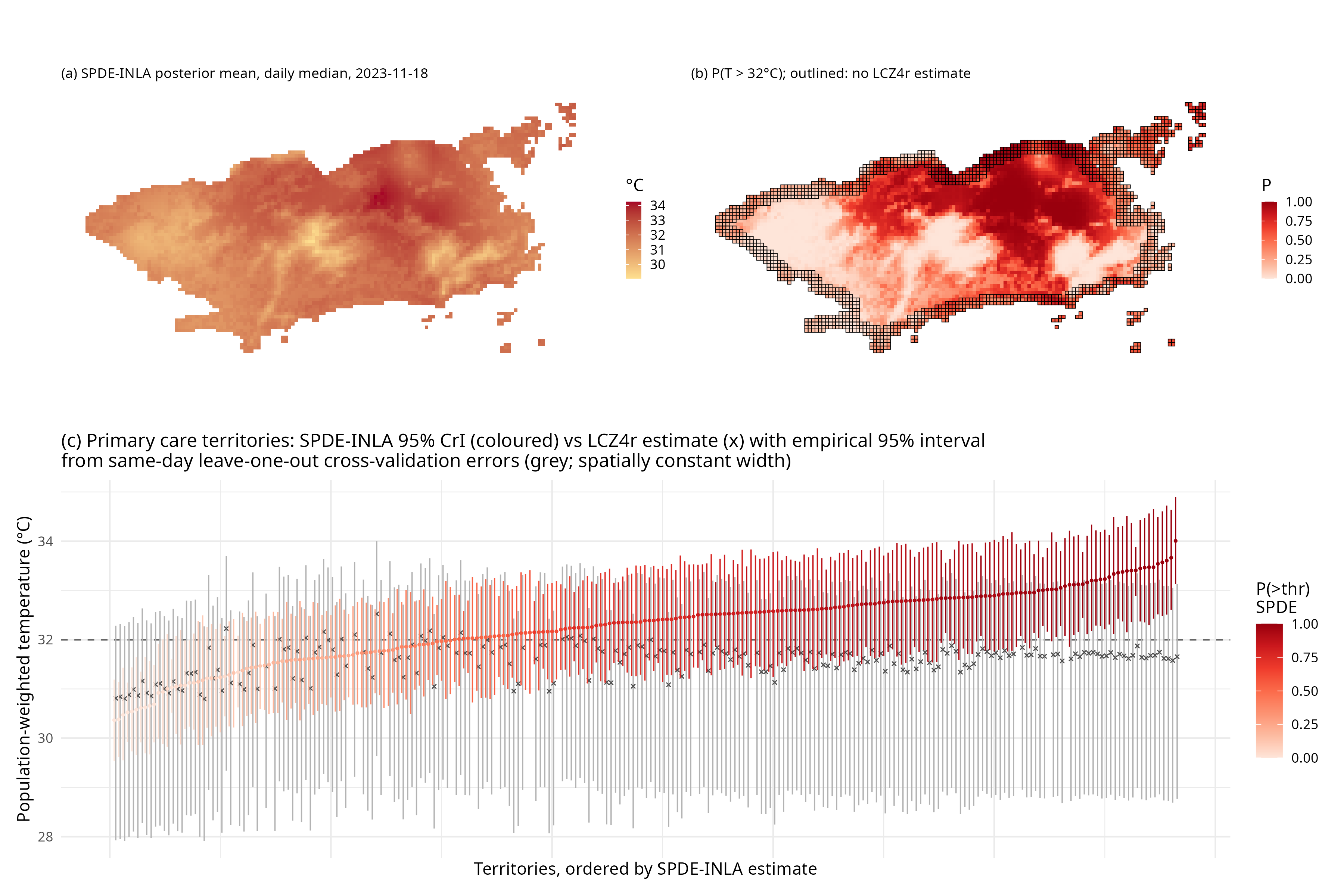}
\caption{Population heat exposure on 18 November 2023. (a) BRUCE posterior mean of daily median temperature. (b) Posterior probability that daily median temperature exceeded 32\textdegree{}C; outlined cells have no LCZ4r estimate (water masking). (c) Population-weighted temperature in each of the 241 primary care territories, ordered by the BRUCE estimate: BRUCE posterior mean and 95\% credible interval (coloured by the probability of exceeding 32\textdegree{}C) and LCZ4r estimate with an empirical 95\% interval derived from same-day leave-one-out cross-validation errors (grey; airport stations excluded; constant width across territories). Dashed line: 32\textdegree{}C threshold.}
\label{fig-exposicao}
\end{figure}

On the hottest day of the study period (18 November 2023), the two methods yielded markedly different assessments of population exposure (Figure~\ref{fig-exposicao}). Under BRUCE, 4.29 million residents (66.7\%; 95\% CrI 2.92--5.13 million) lived in areas where the daily median temperature exceeded 32\textdegree{}C, compared with 0.16 million under the LCZ4r point estimate. In the hottest primary care territories, BRUCE estimates ($\approx$33--34\textdegree{}C) exceeded LCZ4r estimates ($\approx$31.7\textdegree{}C) by 1.5--2\textdegree{}C. BRUCE classified 71 of 241 territories (2.24 million residents) as exceeding the threshold with posterior probability above 0.9 and 25 territories (0.58 million) as below it with probability above 0.9, leaving 145 territories uncertain. The empirical interval derived from LCZ4r cross-validation errors did not allow any territory to be classified with comparable confidence. At the common 500 m resolution, 1.30 million residents lived in areas without an LCZ4r estimate, and 17 territories had no LCZ4r estimate at all.

\section{Discussion}

This study demonstrates that a Bayesian hierarchical spatiotemporal framework can formally fuse four heterogeneous temperature sources (official, ERA5, low-cost IoT, and citizen-science networks) into a single coherent spatial product, with class- and station-level bias explicitly modeled rather than filtered out. This builds on an established tradition of Bayesian hierarchical approaches to combining heterogeneous data sources with differing error structures~\cite{fuentes_model_2005,berrocal_spatio-temporal_2010}, more recently formalized within the INLA-SPDE framework~\cite{villejo_data_2025,zhong_bayesian_2024}, including a parallel application of the same framework to ERA5-linked surface temperature interpolation in a different Brazilian biome~\cite{bianchi_space-time_2025}. Three findings, consistent across the fitting windows examined, support this conclusion. First, the elevation effect was significant and of stable magnitude throughout ($-$0.67 to $-$1.38\textdegree{}C per SD), an identification strengthened by the state-wide extent afforded by the ERA5-Land fusion rather than dependent on data from any single window. Second, decomposing the predicted surface showed that covariates capture the dominant, smooth, city-scale structure of intra-urban temperature, while the spatial random field contributes a smaller but spatially-concentrated correction (25.0\% of variance on average, up to 42.9\% in some windows) that decays with distance from the nearest station and grows on days of greater city-wide thermal amplitude. This pattern indicates that the field is not simply absorbing noise: it tracks genuine, day-specific, hyperlocal departures from the covariate-driven gradient, concentrated precisely where the network can resolve them. Third, the station-level IID effect recovered substantial individual-sensor bias beyond class membership, most frequently among citizen-science stations (76.5\%) but present even among official stations (46.7\%), where it is more plausibly attributable to genuine site-level microclimate than to instrument error, given its reproducibility across independent refits. This finding is consistent with the growing evidence that citizen-weather-station placement itself introduces systematic, non-random bias relative to conventional networks~\cite{sakthivel_spatial_2025}.

Against LCZ4r, BRUCE achieved comparable accuracy, RMSE differences were within bootstrap uncertainty in all three validation windows, but the two methods are better understood as complementary than competitors. LCZ4r's variogram-based interpolation produces smooth, visually coherent surfaces, consistent with its design, but compresses extremes toward each zone's typical value. BRUCE preserves local heterogeneity that two independent lines of evidence indicate is genuine signal rather than noise, provides model-based uncertainty at every location (posterior credible intervals and a predictive distribution of the latent field, evaluated by CRPS).

This distinction has direct consequences for heat-health surveillance. On the hottest day of the study period --- part of the November 2023 heat wave independently documented to have caused approximately 1,392 excess deaths in the municipality~\cite{fernandez-medina_excess_2025} --- the two methods agreed closely at monitoring stations but diverged markedly for the population: 4.29 million residents lived in areas above 32\textdegree{}C under BRUCE, compared with 0.16 million under LCZ4r, a difference driven by the compression of LCZ4r estimates toward the city mean in the warmest territories --- consistent with evidence that heat vulnerability varies by neighborhood microclimate rather than by distance to the nearest monitoring station~\cite{harlan_neighborhood_2006,uejio_intra-urban_2011}, and with recent calls to move heat exposure assessment beyond coarse, area-level estimates toward spatially resolved, uncertainty-aware surfaces capable of informing local-level intervention targeting~\cite{marques_improved_2025,burley_farr_assessing_2026}. BRUCE additionally quantified the probability that each primary care territory exceeded the threshold, allowing 96 of 241 territories to be classified with high confidence (model-based classifications; see Limitations on the scope of the uncertainty intervals), whereas the constant-width interval derived from LCZ4r cross-validation errors did not support confident classification of any territory. Because neither method can be verified at unmonitored locations with the current network, these results show that the choice of interpolation method materially changes who is considered exposed, rather than establishing which method is correct there; independent measurements (e.g., mobile transects or withheld sensors) would be needed for that purpose. This spatial resolution gap is particularly relevant locally: a heat-exposure metric recently developed for Rio de Janeiro's own municipal heat protocol, the Heat Area Above a Threshold (HAAT), was derived and validated using city-wide station averages~\cite{araujo_morais_quantifying_2025}, and a spatially resolved interpolation such as the one developed here could in principle support a territory-specific extension of that protocol.

This framework offers cities facing the same data-sparsity problem as Rio de Janeiro a practical alternative to waiting for a dense, homogeneous official network, since heterogeneous low-cost and citizen-science sources already being collected can be incorporated immediately, with their differential reliability estimated from the data instead of assumed. As the Aurassure and Weather Underground networks continue to grow --- more than doubling the station count available by the final fitting window compared to the first --- the same model structure will accommodate denser input without redesign, and the station-level bias terms identified here can be monitored prospectively to flag sensors that drift or degrade. This operational posture aligns with current WHO guidance on heat--health action plans, which calls for locally resolved exposure information to target interventions rather than relying on city-wide averages~\cite{who_regional_office_for_europe_heathealth_2026}.

These advantages should be read alongside the identifiability limits discussed below: both the spatial range and, in a subset of windows, the field's own temporal correlation remain sensitive to the resolution of the current observation network, and the official-station noise term was fixed rather than estimated to preserve the station-bias identification. Model comparison was deliberately assessed via leave-region-out rather than station-level leave-one-out cross-validation, since holding out individual stations while their spatially-correlated neighbors remain in the training set is known to inflate apparent extrapolation accuracy~\cite{roberts_cross-validation_2017}; the narrower, same-day leave-one-out procedure used to build an empirical error interval for LCZ4r in the population exposure analysis serves a different purpose and is not subject to this concern. None of these choices altered the elevation effect, the study's central and most robust finding, but they bound how finely the spatial field itself should be interpreted at the sub-kilometer scale. Closing this gap --- and eventually linking interpolated exposure to health outcomes at a resolution fine enough to detect intra-urban gradients --- will likely require denser monitoring, satellite-derived land surface temperature as an additional covariate, and independent validation data not used in model fitting.

\section{Limitations}

The spatial range parameter of the Matérn field exhibited substantial sensitivity to prior specification: posterior mean estimates ranged from 1,794m, under a prior anchored at the observation network's median nearest-neighbor distance, to 6,804m, under a weakly-informative alternative prior permitting larger values. This sensitivity was independent of covariate specification, range estimates were essentially unchanged whether elevation and built-up fraction were included or excluded, ruling out confounding between the spatial field and these covariates, and instead reflecting a fundamental limitation of the observation network's spatial resolution for estimating fine-scale range parameters with the currently available station density.

In a subset of fitting windows (10 of 18: seven of nine maximum-temperature windows, and three of nine median-temperature windows --- January, June, and November 2024), the spatial field's temporal correlation parameter (GroupRho) was found to collapse toward values near unity, reproducing, via a distinct mechanism, the same confounding between city-wide and local spatial structure that motivated the introduction of the daily-level term. This collapse was not associated with the presence of the low-cost IoT network (Aurassure): none of the four windows in which Aurassure was operational (June and November 2025, both temperature statistics) exhibited it, while all ten affected windows predate its deployment. GroupRho was consequently fixed (0.55) for these ten windows. A sensitivity analysis confirmed that $\beta_{\mathrm{elev}}$ is robust to the specific GroupRho value chosen for the affected windows, tested against each temperature statistic's own empirically-derived reference value (S1).

Similarly, the observation-level standard deviation for official stations ($\sigma_{\mathrm{ofic}\  }$ = 0.3\textdegree{}C) was fixed a priori rather than estimated, to preserve identifiability of the station-level bias terms ($u_{s}$). An independent empirical check against detrended overnight readings from five official stations (three INMET, two Alerta Rio) yielded a noise estimate of 0.356\textdegree{}C, closely matching the assumed value. However, when $\sigma_{\mathrm{ofic}\  }$  was instead estimated freely in a single-window sensitivity test (November 2025), the model returned a substantially higher posterior mean (0.813\textdegree{}C). This specification choice left the elevation effect ($\beta_{\mathrm{elev}\  }$) unaffected (posterior means differing by $< 0.001 ^{\circ} C$), and individual station-level bias estimates ($u_{s}$) were only modestly attenuated (0--20\% shrinkage in magnitude), with the qualitative pattern of which stations exhibit credible individual bias preserved. We retain $\sigma_{\mathrm{ofic}\  }$ = 0.3\textdegree{}C as the primary specification, consistent with the empirical check, while flagging this residual tension as a target for further investigation.

A block-average treatment of ERA5-Land grid cells, explored as a refinement in an isolated single-window test, produced a markedly improved spatial range estimate with minimal impact on covariate coefficients, but exhibited intermittent numerical instability that could not be reliably resolved within the scope of this study. This refinement is deferred to future work.

We also attempted to extend the model with a topographic barrier term (non-Euclidean Matérn field respecting coastline and massif boundaries), motivated by Rio de Janeiro's fragmented topography. This extension proved structurally incompatible with the temporal group/AR1 replication scheme used throughout this study, no publicly available barrier implementation currently supports grouped/replicated fields, and a simplified, temporally-unstructured single-day pilot fit produced implausible spatial-field extrapolation beyond the study domain.

\subsection{Region-specific extrapolation}

Leave-region-out cross-validation errors did not exhibit a consistent relationship with distance to the nearest training-set station across regions (Pearson r ranging from $-$0.90 to +0.92 depending on the held-out region), suggesting that regional-level prediction difficulty reflects unmodeled local spatial heterogeneity rather than a simple distance-decay effect.

\subsection{Airport effect}

The airport class effect (+1.1 to +2.4 \textdegree{}C across folds) is shared by all airports, whereas individual airports differ from it by about 1--1.5 \textdegree{}C on average (e.g., SBJR +1.5 \textdegree{}C and SBGL $-$1.1 \textdegree{}C, in opposite directions; S1); because the station-specific effect $u_{s}$ of a held-out airport is unavailable, held-out airport predictions retain this residual error.

\subsection{Covariate selection}

The covariate set was fixed a priori on physical grounds and compared with alternatives in three median windows only (S1). Elevation was essential, and the SVF was retained for theoretical reasons without a detectable gain in model fit; a parallel comparison for the daily minimum, where its effect is expected to be strongest, showed a negative coefficient in all three windows but likewise no improvement in fit. Distance to the coast showed a consistent, credible coefficient but an inconsistent gain in marginal likelihood and no gain in station-level predictive score; it is a candidate for future versions.

\subsection{Comparison with LCZ4r}

The comparison with LCZ4r was restricted to the common set of held-out station-days for which LCZ4r produces a valid estimate. Coastal and water masking removes all of Planning Area 1 and several stations elsewhere, so the comparison does not cover the most coastal areas, where only BRUCE yields predictions, and it rests on three windows.

\subsection{Scope of the uncertainty intervals}

Credible intervals and the CRPS reported here refer to the latent temperature field and do not include the station-specific effect ($\sigma_{u}$ $\approx$ 0.8--1.0 \textdegree{}C) or the measurement error of a new sensor; they should not be read as prediction intervals for the reading of an unmonitored sensor. Accordingly, nominal 90\% latent-field intervals covered about half of held-out sensor readings in January 2025 (51\%) and about one third in June and November 2025 (32\% and 33\%), and classifications based on posterior exceedance probabilities should be read as conditional on the model.

\section*{Author contribution}

\textbf{AG}: Conceptualization, Data curation, Formal analysis, Investigation, Methodology, Project administration, Software, Supervision, Validation, Visualization, Writing -- original draft, Writing -- review \& editing. \textbf{RO}, \textbf{LM}, \textbf{FV}, \textbf{JM}, \textbf{VS}, \textbf{GM}, \textbf{CF}: Writing -- original draft, Writing -- review \& editing. \textbf{OC}: Conceptualization, Investigation, Writing -- review \& editing.

% \section*{Acknowledgments}

% We thank the team at the Centro de Inteligência Epidemiológica (CIE; Epidemiological Intelligence Center), Rio de Janeiro City Hall, for the technical and scientific discussions on the health impacts of heat across the city.

\section*{Funding}

Coordenação de Aperfeiçoamento de Pessoal de Nível Superior (CAPES) for the doctoral scholarship granted to AG.

\section*{Data and code availability}

The analysis code is available at https://github.com/asgouveiaa/BRUCE-paper. The station data used in this study include observations from a privately operated sensor network and cannot be shared publicly; the processed data underlying the results are available from the corresponding author upon reasonable request.

\section*{Declaration of generative AI and AI-assisted technologies in the manuscript preparation process}

During the preparation of this work the authors used Claude (Anthropic) to improve the language and structure of the general text, and to assist in debugging analysis code (R scripts for the supplementary analyses and tables). After this, the authors reviewed and edited the content and take full responsibility for the content of the published article.

\bibliographystyle{elsarticle-num}

\end{document}

\maketitle

\section{GroupRho fixation: rationale and validation}

GroupRho, the temporal autocorrelation parameter of the spatial field's AR1 group structure, exhibited weak identifiability in 10 of 18 fitting windows (Table~\ref{tab:grouprho-interventions}), converging toward the boundary of its parameter space (GroupRho$\rightarrow$1) or failing to converge entirely. Following the same pragmatic approach already applied to the daily-level AR1 precision, GroupRho was fixed at 0.55 for these windows.

This value was validated via sensitivity analysis against each statistic's own empirically-derived reference: for median temperature, the mean GroupRho across six freely-estimated windows was 0.70 (range 0.64--0.80); for maximum temperature, the mean across two freely-converged windows was 0.40 (0.397, 0.407) --- a lower value consistent with maximum temperature's greater sensitivity to transient, sub-daily conditions. Refitting affected windows with these empirically-derived alternative values instead of 0.55 changed $\beta_{\mathrm{elev}}$ by 0.01 and 0.03 \textdegree{}C (Table~\ref{tab:grouprho-sensitivity}), confirming that this pragmatic constraint does not materially bias the covariate of primary scientific interest.

\begin{table}[htbp]
\centering
\small
\setlength{\tabcolsep}{4pt}
\caption{Windows requiring GroupRho intervention, out of 18 total (9 windows $\times$ 2 temperature statistics). ``Collapsed'' windows converged but with GroupRho at the boundary of its parameter space; ``non-convergent'' windows failed to converge under free estimation across repeated attempts.}
\label{tab:grouprho-interventions}
\resizebox{\ifdim\width>\linewidth\linewidth\else\width\fi}{!}{%
\begin{tabular}{lcccc}
\toprule
Window & Variable & Initial state & GroupRho estimate & Resolution \\
\midrule
Jan 2024 & Median & Collapsed (GroupRho$\rightarrow$1) & --- & Fixed at 0.55 \\
Jan 2023 & Max & Collapsed (GroupRho$\rightarrow$1) & --- & Fixed at 0.55 \\
Jun 2023 & Max & Collapsed (GroupRho$\rightarrow$1) & --- & Fixed at 0.55 \\
Nov 2023 & Max & Collapsed (GroupRho$\rightarrow$1) & --- & Fixed at 0.55 \\
Jan 2024 & Max & Collapsed (GroupRho$\rightarrow$1) & --- & Fixed at 0.55 \\
Jun 2024 & Max & Collapsed (GroupRho$\rightarrow$1) & --- & Fixed at 0.55 \\
Nov 2024 & Max & Collapsed (GroupRho$\rightarrow$1) & --- & Fixed at 0.55 \\
Jan 2025 & Max & Collapsed (GroupRho$\rightarrow$1) & --- & Fixed at 0.55 \\
Jun 2024 & Median & Non-convergent (repeated crash) & --- & Fixed at 0.55 from initialization \\
Nov 2024 & Median & Non-convergent (repeated crash) & --- & Fixed at 0.55 from initialization \\
\bottomrule
\end{tabular}%
}
\end{table}

\begin{table}[htbp]
\centering
\small
\setlength{\tabcolsep}{4pt}
\caption{Sensitivity of the elevation coefficient to the fixed GroupRho value, tested against each temperature statistic's own empirically-derived reference value from freely-converged windows. Differences (0.01\textdegree{}C for median, 0.03\textdegree{}C for max) are well within one posterior standard deviation, confirming that $\beta_{\mathrm{elev}}$ is not materially sensitive to the specific value chosen.}
\label{tab:grouprho-sensitivity}
\resizebox{\ifdim\width>\linewidth\linewidth\else\width\fi}{!}{%
\begin{tabular}{lccc}
\toprule
Test & GroupRho value & $\beta_{\mathrm{elev}}$ mean & $\beta_{\mathrm{elev}}$ SD \\
\midrule
\multirow{2}{*}{Median (Jan 2024)} & 0.70 (empirical, healthy median windows) & $-$1.29 & 0.032 \\
 & 0.55 (used) & $-$1.28 & 0.032 \\
\multirow{2}{*}{Max (Nov 2023)} & 0.40 (empirical, healthy max windows) & $-$0.976 & 0.057 \\
 & 0.55 (used) & $-$0.942 & 0.072 \\
\bottomrule
\end{tabular}%
}
\end{table}

\section{Empirical validation of the fixed official-station observation noise}

Unlike the low-cost IoT (Aurassure) and citizen-science (Weather Underground) networks, whose observation-level precisions are freely estimated within the model, the official-station observation noise is fixed a priori ($\sigma_{\mathrm{ofic}}$ = 0.3\textdegree{}C), reflecting an assumption rather than a documented instrument specification. We evaluated this assumption with an independent empirical estimate of instrument-level noise, external to the hierarchical spatiotemporal model.

Rationale. Estimating $\sigma_{\mathrm{ofic}}$ by simply freeing it within the full model is confounded: because the model already includes a station-level random effect ($u_{s}$) intended to absorb persistent site-specific bias (e.g., microclimatic exposure, siting artifacts), a freely estimated $\sigma_{\mathrm{ofic}}$ can instead absorb genuine site heterogeneity that is not true measurement noise, inflating the noise estimate. Indeed, a sensitivity test in which $\sigma_{\mathrm{ofic}}$ was estimated freely within the full model produced $\sigma_{\mathrm{ofic}\  }^{h}$ = 0.813\textdegree{}C --- larger than the freely estimated precisions for both the Aurassure (0.686\textdegree{}C) and Weather Underground (0.576\textdegree{}C) networks, a physically implausible ordering given that official stations use professional-grade calibrated instruments, and one that materially shifted the elevation coefficient ($\Delta$ = 0.311\textdegree{}C), a covariate of primary scientific interest. This motivated an independent, model-external validation.

Method. We used raw hourly observations from a subset of five official stations spanning both official sub-networks present in the dataset: three INMET automatic stations (A602, A621, A636) and two Alerta Rio municipal stations (19 --- Barra/Riocentro; 28 --- Alto da Boa Vista). Data were restricted to overnight hours (00:00--06:00 local time), when near-surface temperature typically evolves slowly and smoothly, minimizing genuine short-timescale thermal variability. For each station-night with at least three hourly readings, a polynomial trend in time-of-night (linear for 3--4 points, quadratic for 5 or more) was fit by ordinary least squares, and residuals from this fit were retained as an estimate of point-to-point measurement noise, net of the night's genuine (smooth) thermal trajectory. Residuals were pooled across all station-nights (both overall and by sub-network) and their standard deviation compared against the assumed $\sigma_{\mathrm{ofic}}$.

Results. The pooled residual standard deviation was 0.356\textdegree{}C overall, closely matching the assumed value (ratio: 1.19$\times$). Sub-network estimates were nearly identical --- 0.352\textdegree{}C (INMET) and 0.357\textdegree{}C (Alerta Rio) --- despite substantial differences in instrumentation between the two networks, supporting the robustness of the estimate rather than it being an artifact of a single station or network type. Because within-night residual meteorological variability not captured by the polynomial detrend (e.g., wind gusts, transient cloud cover) would inflate rather than deflate this estimate, 0.356\textdegree{}C is interpreted as a conservative upper bound on pure instrument noise --- if anything, reinforcing rather than undermining the adequacy of the assumed 0.3\textdegree{}C value.

Conclusion. This independent, model-external estimate is consistent with the value assumed in the main analysis ($\sigma_{\mathrm{ofic}}$ = 0.3\textdegree{}C) and inconsistent with the substantially larger value recovered when $\sigma_{\mathrm{ofic}}$ is freely estimated within the full hierarchical model, corroborating the interpretation that the latter conflates measurement noise with genuine station-level bias. We therefore retain $\sigma_{\mathrm{ofic}}$ fixed at 0.3\textdegree{}C throughout the main analysis.

\begin{table}[htbp]
\centering
\small
\setlength{\tabcolsep}{4pt}
\caption{Station-level empirical noise estimates for official stations, obtained as the residual standard deviation after per-night polynomial detrending of hourly overnight readings (00:00--06:00 local time). The methodology is described above.}
\label{tab:sigma-ofic-validation}
\resizebox{\ifdim\width>\linewidth\linewidth\else\width\fi}{!}{%
\begin{tabular}{llr}
\toprule
\textbf{Station ID} & \textbf{Network} & \textbf{Residual SD (\textdegree{}C)} \\
\midrule
A602, A621, A636 & INMET (automatic) & 0.352 \\
19, 28 & Alerta Rio (municipal) & 0.357 \\
\midrule
\textbf{Pooled (all 5 stations)} &  & \textbf{0.356} \\
\bottomrule
\end{tabular}%
}
\end{table}

\section{Verification of the airport class-level / station-level bias decomposition}

During preliminary model development, a class-level fixed effect for airport weather stations ($\beta_{\mathrm{aeroporto}}$) was found to exhibit heterogeneous bias in opposite directions across individual airport stations. This raised the question of whether the class-level term should be retained once the station-level IID effect ($u_{s}$) was introduced later in model development, since $u_{s}$ could in principle absorb exactly this kind of station-specific heterogeneity, making the earlier concern obsolete rather than making the class effect invalid. This note verifies that empirically, using the already-fitted leave-region-out models rather than a new model fit.

For each of the 15 leave-region-out refits (5 regions $\times$ 3 months: January, June, November 2025), the saved INLA fit object was inspected directly (no refitting). Two quantities were extracted: the posterior of the class-level fixed effect ($\beta_{\mathrm{aeroporto}}$), and the posterior station-level IID effect ($u_{s}$) restricted to the five REDEMET airport stations (SBRJ, SBGL, SBSC, SBJR, SBAF) present in that fold's training set.

$\beta_{\mathrm{aeroporto}}$ was positive in all 15 refits (posterior mean +1.1 to +2.4 \textdegree{}C), consistent with a genuine warm offset for paved, low-vegetation airport surfaces. Conditional on $\beta_{\mathrm{aeroporto}}$, the station-level effect $u_{s}$ of individual airports was stable across the 12 refits in which each airport was retained in training: SBAF +0.92 \textdegree{}C, SBJR +1.46 \textdegree{}C, SBGL $-$1.06 \textdegree{}C and SBSC $-$1.08 \textdegree{}C (mean across refits), each with the same sign in every refit (Table~\ref{tab:aeroporto-us-decomposition}), whereas SBRJ was unstable (mean +0.08 \textdegree{}C, sign varying).

\begin{table}[htbp]
\centering
\small
\setlength{\tabcolsep}{4pt}
\caption{Station-level effect $u_{s}$ of the five REDEMET airport stations across the 12 leave-region-out refits in which each station was retained in the training set, conditional on the class-level airport effect $\beta_{\mathrm{aeroporto}}$. Four stations kept a stable sign in all refits; SBRJ did not.}
\label{tab:aeroporto-us-decomposition}
\resizebox{\ifdim\width>\linewidth\linewidth\else\width\fi}{!}{%
\begin{tabular}{lcccc}
\toprule
\textbf{Station} & \textbf{Mean $u_{s}$ (\textdegree{}C)} & \textbf{Range across refits (\textdegree{}C)} & \textbf{Refits} & \textbf{Sign} \\
\midrule
SBJR & +1.46 & +0.85 to +1.86 & 12 & positive in 12 of 12 \\
SBAF & +0.92 & +0.50 to +1.56 & 12 & positive in 12 of 12 \\
SBGL & $-$1.06 & $-$1.56 to $-$0.47 & 12 & negative in 12 of 12 \\
SBSC & $-$1.08 & $-$1.69 to $-$0.50 & 12 & negative in 12 of 12 \\
SBRJ & +0.08 & $-$0.78 to +1.12 & 12 & varies \\
\bottomrule
\end{tabular}%
}
\end{table}

This confirms the two terms are complementary rather than redundant: $\beta_{\mathrm{aeroporto}}$ captures the shared, physically plausible offset for the airport class as a whole; $u_{s}$ captures the residual, station-specific deviation from that class mean, in either direction. The opposite-direction heterogeneity that originally motivated concern about the class-level term is fully attributable to station-specific siting differences (e.g., proximity to pavement vs. grass, shading, exposure), not to a deficiency of the class-level correction itself. Both terms are therefore retained jointly in the final model.

\section{Case investigation --- 8 November 2025}

The largest daily errors in November 2025 were examined for a single dominant event. 8 November 2025 had the second-highest daily RMSE of the window (2.36 \textdegree{}C), but excluding it changed the November leave-region-out RMSE only from 1.66 to 1.63 \textdegree{}C; the November result is therefore not driven by a single anomalous day.

\section{Comparison against reference interpolation methods}

\begin{table}[htbp]
\centering
\small
\setlength{\tabcolsep}{4pt}
\caption{Root mean square error (\textdegree{}C) under leave-region-out cross-validation. The common set comprises the held-out station-days for which LCZ4r produces a valid estimate (coastal/water masking removes all of Planning Area 1 and several stations elsewhere); predictions at airport stations include the class-level airport effect. Difference = BRUCE $-$ LCZ4r (negative favours BRUCE), with a 95\% station-level bootstrap interval (2,000 replicates). Bold = lowest RMSE.}
\label{tab:rmse-comparison-supp}
\resizebox{\ifdim\width>\linewidth\linewidth\else\width\fi}{!}{%
\begin{tabular}{llcccc}
\toprule
\textbf{Station set} & \textbf{Month} & \textbf{n} & \textbf{BRUCE} & \textbf{LCZ4r} & \textbf{Difference (95\% CI)} \\
\midrule
Common set, with airports & Jan 2025 & 302 & \textbf{1.90} & 2.21 & $-$0.31 ($-$0.87; 0.27) \\
 & Jun 2025 & 285 & 2.13 & \textbf{2.08} & +0.05 ($-$0.77; 0.94) \\
 & Nov 2025 & 276 & \textbf{1.76} & 1.83 & $-$0.07 ($-$0.60; 0.52) \\
Common set, no airports & Jan 2025 & 210 & \textbf{1.69} & 1.88 & -- \\
 & Jun 2025 & 196 & 2.12 & 2.12 & -- \\
 & Nov 2025 & 186 & \textbf{1.52} & 1.71 & -- \\
Full network (BRUCE only) & Jan 2025 & 457 & 1.82 & -- & -- \\
 & Jun 2025 & 431 & 2.20 & -- & -- \\
 & Nov 2025 & 426 & 1.66 & -- & -- \\
\bottomrule
\end{tabular}%
}
\end{table}

\begin{table}[htbp]
\centering
\small
\setlength{\tabcolsep}{4pt}
\caption{Root mean square error (\textdegree{}C) by Planning Area and month under leave-region-out cross-validation (n = station-days; Planning Area 3 in November has only 6). LCZ4r is evaluated on the common set of station-days where it produces a valid estimate; where it produces none (coastal/water masking, all of Planning Area 1), BRUCE is evaluated on all held-out station-days of the region. Bold = lowest RMSE. Difference = BRUCE $-$ LCZ4r (negative favours BRUCE).}
\label{tab:rmse-region-supp}
\resizebox{\ifdim\width>\linewidth\linewidth\else\width\fi}{!}{%
\begin{tabular}{lccccc}
\toprule
\textbf{Month} & \textbf{Planning Area} & \textbf{n} & \textbf{BRUCE} & \textbf{LCZ4r} & \textbf{Difference} \\
\midrule
Jan 2025 & 1 & 62 & 2.10 & -- & -- \\
Jan 2025 & 2 & 62 & \textbf{0.98} & 2.17 & $-$1.19 \\
Jan 2025 & 3 & 31 & 1.89 & \textbf{0.79} & +1.09 \\
Jan 2025 & 4 & 86 & \textbf{2.69} & 2.93 & $-$0.24 \\
Jan 2025 & 5 & 123 & \textbf{1.56} & 1.84 & $-$0.29 \\
Jun 2025 & 1 & 58 & 1.89 & -- & -- \\
Jun 2025 & 2 & 58 & \textbf{1.94} & 2.56 & $-$0.63 \\
Jun 2025 & 3 & 29 & 2.52 & \textbf{0.79} & +1.73 \\
Jun 2025 & 4 & 89 & \textbf{2.31} & 2.38 & $-$0.07 \\
Jun 2025 & 5 & 109 & 1.95 & \textbf{1.75} & +0.20 \\
Nov 2025 & 1 & 60 & 1.63 & -- & -- \\
Nov 2025 & 2 & 60 & \textbf{1.56} & 2.46 & $-$0.90 \\
Nov 2025 & 3 & 6 & 2.51 & \textbf{1.36} & +1.15 \\
Nov 2025 & 4 & 90 & 2.20 & \textbf{2.18} & +0.02 \\
Nov 2025 & 5 & 120 & 1.40 & \textbf{1.02} & +0.38 \\
\bottomrule
\end{tabular}%
}
\end{table}

\begin{table}[htbp]
\centering
\small
\setlength{\tabcolsep}{4pt}
\caption{Continuous ranked probability score (CRPS) for BRUCE under leave-region-out cross-validation, by month (overall mean 1.32\textdegree{}C). The predictive distribution is that of the latent temperature field (fixed effects, daily term and spatial field); it excludes the station-specific effect, the measurement error of a new sensor and, for airport stations, the class-level airport effect. Reported for BRUCE only; LCZ4r's deterministic variogram-based interpolation does not produce a comparable predictive distribution.}
\label{tab:crps-supp}
\resizebox{\ifdim\width>\linewidth\linewidth\else\width\fi}{!}{%
\begin{tabular}{lcc}
\toprule
Month & n & CRPS (\textdegree{}C) \\
\midrule
January 2025 & 457 & 1.16 \\
June 2025 & 431 & 1.65 \\
November 2025 & 426 & 1.16 \\
\bottomrule
\end{tabular}%
}
\end{table}

\section{Evidence that BRUCE's local heterogeneity reflects genuine signal}

Two lines of evidence support this. First, on the common set of held-out station-days, BRUCE's predictions correlated more strongly with observations than LCZ4r's in all three windows (January: r = 0.84 vs. 0.69; June: 0.79 vs. 0.72; November: 0.69 vs. 0.60). Much of this correlation reflects the city-wide daily cycle shared by all stations; after removing each day's mean across stations, the within-day correlation was 0.57 vs. $-$0.02 (January), 0.54 vs. 0.12 (June) and 0.41 vs. 0.07 (November), showing that BRUCE, unlike LCZ4r, captures the spatial pattern across stations within a day. The advantage persisted when airport stations were excluded (r = 0.89 vs. 0.74, 0.80 vs. 0.72 and 0.73 vs. 0.66; within-day 0.76 vs. 0.06, 0.69 vs. 0.23 and 0.38 vs. 0.12). On the full network, where LCZ4r has no valid estimate in part of the domain, BRUCE's correlations were 0.77, 0.71 and 0.65 (within-day 0.34, 0.31 and 0.31). Second, the station-level effects $u_{s}$ of the airports showed a stable pattern across the 12 refits in which each was retained in training (SBJR and SBAF consistently warmer, SBGL and SBSC consistently cooler than the airport-class mean; SBRJ unstable; Table~\ref{tab:aeroporto-us-decomposition}), indicating reproducible site-specific signal rather than overfit measurement noise.

\section{Covariate selection (model comparison)}

We also evaluate the median temperature result in three windows (June 2023, January 2025 and November 2025) with nested and alternative covariate sets, keeping the data, mesh, priors, likelihood structure and hyperparameter settings of the main analysis. Candidate additions were distance to the coast, distance to inland water bodies and mean building height (GHSL GHS-BUILT-H, 2018), all standardised as the other covariates. Models were compared with the log marginal likelihood (mlik), WAIC and the log predictive score (LPML, computed on station observations only, on rows common to all models). WAIC and LPML are in-sample criteria and should be read as supportive rather than as out-of-sample validation. Differences in Table~\ref{tab-abl} are relative to the model used in the paper (positive $\Delta$mlik and $\Delta$LPML, and negative $\Delta$WAIC, favour the alternative); values are summed over the three windows.

\begin{table}[htbp]
\centering
\small
\setlength{\tabcolsep}{4pt}
\caption{Covariate comparison relative to the model used in the paper, summed over three median windows (June 2023, January 2025, November 2025).}
\label{tab-abl}
\resizebox{\ifdim\width>\linewidth\linewidth\else\width\fi}{!}{%
\begin{tabular}{lccccc}
\toprule
\textbf{Model} & \textbf{$\Delta$ mlik} & \textbf{$\Delta$ WAIC} & \textbf{$\Delta$ LPML} & \textbf{Coef. (\textdegree{}C per SD)} & \textbf{Windows with CrI $\neq$ 0} \\
\midrule
No covariates & $-$237.3 & +240.5 & $-$188.4 & -- & -- \\
Elevation only & +2.9 & +13.3 & $-$2.0 & -- & -- \\
Elevation + built-up & +4.9 & $-$1.0 & +0.6 & -- & -- \\
\textbf{Paper (elevation + built-up + SVF)} & \textbf{0} & \textbf{0} & \textbf{0} & -- & -- \\
Paper $-$ elevation & $-$234.5 & +215.8 & $-$182.4 & -- & -- \\
Paper $-$ built-up & $-$2.4 & +11.7 & $-$1.5 & -- & -- \\
Paper + distance to water & $-$9.9 & $-$7.8 & +1.7 & +0.03 & 0 / 3 \\
Paper + distance to coast & +7.9 & $-$152.3 & $-$4.5 & $-$0.20 & 3 / 3 \\
Paper + building height & $-$8.5 & $-$5.4 & 0.0 & +0.06 & 0 / 3 \\
Paper + all three & $-$8.0 & $-$161.3 & $-$3.7 & -- & -- \\
\bottomrule
\end{tabular}%
}
\end{table}

Elevation is indispensable, since removing it lowered the marginal likelihood by 109, 83 and 43 log units in the three windows. Built-up fraction contributes modestly (removal changed mlik by $-$4.3, $-$1.0 and +2.9). The SVF coefficient was not distinguishable from zero in any window (posterior means +0.34, $-$0.32 and $-$0.03 \textdegree{}C per SD, all credible intervals including zero), and the model without SVF had equal or higher marginal likelihood in all three windows (+1.2, +0.9 and +2.7) with indistinguishable WAIC. Distance to water and building height lowered the marginal likelihood in all three windows ($-$3.6, $-$3.3, $-$3.0 and $-$2.0, $-$2.6, $-$3.9) and their coefficients were never credibly different from zero. Distance to the coast was the only candidate with some support: its coefficient was negative and credible in all three windows (mean $-$0.20 \textdegree{}C per SD), but the marginal-likelihood gain was concentrated in June 2023 (+7.0, against +0.0 in January 2025 and +0.9 in November 2025) and the station-level LPML did not improve ($-$4.5). The large WAIC gain ($-$152) is driven by the much larger number of ERA5 observations and should not be read as improved prediction at stations. We therefore kept the three-covariate specification for parsimony and note distance to the coast as the most promising extension.

The SVF was therefore retained on theoretical grounds, because it is part of the physical model of urban canopy heat storage and is expected to matter mainly for nocturnal minimum temperature. To assess this effect, we refitted the same three windows for the daily minimum temperature. The SVF coefficient was negative in all three windows (mean $-$0.26 \textdegree{}C per SD; credible interval excluding zero in one), consistent with greater nocturnal radiative cooling under open sky, whereas for the daily median it was close to zero (mean $-$0.001 \textdegree{}C per SD). However, including the SVF did not improve the marginal likelihood, WAIC or predictive score for the minimum either ($\Delta$ log marginal likelihood $-$2.9 summed over windows). The evidence therefore supports the expected direction of the effect but not a detectable improvement in fit, and the SVF was retained on theoretical grounds.